# Architectural Tactics for Big Data Cybersecurity Analytic Systems: A Review

Faheem Ullah[a, b], Muhammad Ali Babar[a, b]

[a]Cyber Security Adelaide, University of Adelaide, Australia
[b]CREST- the Centre for Research on Engineering Software Technologies, Australia

**Abstract**

*Context:* Big Data Cybersecurity Analytics is increasingly becoming an important area of research and practice aimed at protecting networks, computers, and data from unauthorized access by analysing security event data using big data tools and technologies. Whilst a plethora of Big Data Cybersecurity Analytic Systems have been reported in the literature, there is a lack of a systematic and comprehensive review of the literature from an architectural perspective. *Objective:* This paper reports a systematic review aimed at identifying the most frequently reported quality attributes and architectural tactics for Big Data Cybersecurity Analytic Systems. *Method:* We used Systematic Literature Review (SLR) method for reviewing 74 primary studies selected using well-defined criteria. *Results:* Our findings are twofold: (i) identification of 12 most frequently reported quality attributes and the justification for their significance for Big Data Cybersecurity Analytic Systems; and (ii) identification and codification of 17 architectural tactics for addressing the quality attributes that are commonly associated with Big Data Cybersecurity Analytic systems. The identified tactics include six performance tactics, four accuracy tactics, two scalability tactics, three reliability tactics, and one security and usability tactic each. *Conclusion:* Our findings have revealed that (a) despite the significance of interoperability, modifiability, adaptability, generality, stealthiness, and privacy assurance, these quality attributes lack explicit architectural support in the literature (b) empirical investigation is required to evaluate the impact of codified architectural tactics (c) a good deal of research effort should be invested to explore the trade-offs and dependencies among the identified tactics (d) there is a general lack of effective collaboration between academia and industry for supporting the field of Big Data Cybersecurity Analytic Systems and (e) more research is required on the comparative analysis among big data processing frameworks (i.e., Hadoop, Spark, and Storm) when used for Big Data Cybersecurity Analytic Systems.

## 1. Introduction

Cybersecurity is a set of tools, practices, and guidelines that can be used to protect computer networks, software programs, and data from attack, damage, or unauthorized access [1]. Cybersecurity is critical from two perspectives: 1) we live in a society that is digitally hyper-connected for social engagements, businesses, education, healthcare, and many other aspects of everyday lives; 2) the increasing dependence on digitalization motivates the advancement of threat landscape where minor and major (e.g., Advanced Persistent Threats (APT)) cyber security attacks are getting more organized and sophisticated day-by-day. According to a recent study [2], there can be on average 27.4% more cyber attacks for an organization in 2017 compared to 2016. The accelerating cyber attacks are quite detrimental and costing organizations on average $11.7 million per year.

The persistent rise in the successful execution of cyber attacks and its widespread effects show that the traditional cybersecurity tools and practices are not able to cope with the sophisticated threat landscape. Cardenas et al. [3] report the reasons for such inadequacy of traditional cybersecurity that include (i) retaining a large amount of data (ii) analyzing unstructured data (iii) managing large data warehouses (iv) responding in real-time and (v) detecting Advanced Persistent Threats (APT). To address these limitations, Cardenas et al. [4] propose an evolution model for cybersecurity that encourages the incorporation of big data tools and technologies. There exist hundreds of such tools and technologies and are well-documented in the academic literature [5]. Some of the prominent big data tools include Hadoop, Spark, Storm, Flume, Hbase, Hive, Kafka, Cassandra, and Mahout. It has been proposed in [4] that big data tools and technologies would transform cybersecurity analytics by enabling organizations to (i) collect a large amount of security-related heterogeneous data from diverse sources such as networks, databases, and applications (ii) perform deep security analytics at real-time and (iii) provide a consolidated view of the security-related information.

This combination of cybersecurity solutions and big data tools has given birth to the term 'Big Data Cybersecurity Analytic systems', which refers to systems that collect large amount of security event data from different sources and analyze it using big data tools and technologies for detecting attacks either through attack pattern matching or identifying anomalies. Security events are entities of information that contain valuable insight pertinent to the cybersecurity of an organization. The sources from which security event data is collected include, but not limited to, network traffic data, firewall logs, web logs, system logs, router access logs, database access logs, and application logs. For instance, IP addresses and their cyber activities directed towards an organization are analyzed by the Big Data Cybersecurity Analytic System of an organization. In case the analysis results indicate that a particular IP is involved in incomplete or failed connections that are beyond a certain threshold, the IP is considered malicious and further action is taken against the IP address holder. In some cases, the Big Data Cybersecurity Analytic Systems are generic and applicable for a variety of attacks (e.g., intrusion detection system or alert correlator); in other cases, such systems are applicable to specific attacks (e.g., phishing detection and malware detection).

The cybersecurity analytic systems (e.g., intrusion detection system) are implemented both with big data tools and without big data tools. For example, several studies report intrusion detection systems that does not incorporate any big data tools. On the other hand, several studies report intrusion detection systems that incorporate big data tools. This review is only focussed on the cybersecurity analytic systems (both generic and specific) that incorporate big data tools and technologies. Since a number of such tools and technologies exists, therefore, to precisely guide the scope of this study, we decided that any cybersecurity system that does not leverage big data processing framework (e.g., Hadoop, Spark, and Storm) was out of the scope of this study. For the sake of brevity, we will use the term *security analytics* instead of *Big Data Cybersecurity Analytics* throughout the paper. Similarly, we will use the term *security analytic system* to refer to *Big Data Cybersecurity Analytic System*.

Although the application of big data tools and technologies in cybersecurity domain offers significant promises, it presents some challenges that are unique to cybersecurity as compared to big data analytics in other domains such as healthcare, transportation, entertainment, and banking. The challenges have been highlighted in [3] and [6] which include (i) privacy assurance (ii) authenticity and integrity of security event data (iii) lack of datasets (iv) asymmetrical cost of misclassification (v) adversarial learning and (vi) attack time scale. These challenges make the architecture of a security analytic system both complex and critical [4, 7]. Therefore, software architects need to carefully architect a security analytic system that establishes the required balance among various quality concerns (e.g., performance, accuracy, or reliability) [8]. The architectural challenges and increasing importance of security analytics motivated researchers to investigate different solutions for security analytics.

We noticed that there is no systematic effort to review the state-of-the-art of security analytics from an architectural perspective. By architectural perspective, we mean the identification and justification of architecturally significant requirements (i.e., quality attributes) and how these requirements have been addressed (i.e., architectural tactics [9]). There is an increasing realisation of the importance of systematically reviewing and understanding the architectural aspects of the state-of-the-art systems for building and leveraging evidence-based architectural knowledge in different domains such as cyber-foraging [10], energy efficiency in cloud [11], ambient assisted living systems [12], and self-adaptive systems [13]. Our effort is similar to these works but in a different area of research and practice, i.e., security analytic systems. We systematically selected and rigorously reviewed 74 relevant papers on the topic. Our analysis and synthesis of the data extracted from the reviewed papers have enabled us to make the following key contributions:

1. An identification of the most important quality attributes for security analytic systems;
2. A catalogue of 17 architectural tactics for addressing quality concerns in security analytic systems;
3. An identification of the potential areas of research to advance the state-of-the-art on architecting big data security analytic systems.

The rest of this paper is organized as follows. Section 2 presents the research methodology followed for this SLR. Section 3 reports the demographic view of the reviewed primary studies. Section 4 presents the identified critical quality attributes, which is followed by Section 5 that presents the codified architectural tactics. Section 6 provides the discussion on the main findings and identifies potential areas for future research. Section 7 reports the related work. Section 8 discusses the limitations of this SLR. Finally, Section 9 concludes the paper.

## 2. Research methodology

We used Systematic Literature Review (SLR) [14] method for conducting this review. An SLR is aimed at systematically identifying and selecting relevant studies to be reviewed on a particular topic and rigorously analysing and synthesizing the extracted data from the reviewed studies to answer a set of research questions. We followed the SLR guidelines reported in [14]. The main components of our review protocol were: (i) research questions; (ii) search strategy; (iii) inclusion and exclusion criteria; (iv) study selection; and (v) data extraction and synthesis procedures. These components are discussed in the following subsections.

2.1. Research questions

This study is aimed at identifying and summarizing the most important quality attributes (RQ1) and architectural tactics (RQ2) for achieving those quality attributes of security analytic systems. Table 1 presents the research questions and their respective motivators for this SLR.

Table 1. Research questions for this SLR.

| Research question | Motivation |
| --- | --- |
| **RQ1:** Which are the most important quality attributes for security analytic systems? | To find out the most frequently emphasized quality attributes and the motivation behind their importance for a security analytic system. |
| **RQ2:** What are the architectural tactics for addressing quality concerns in security analytic systems? | To gain a detailed understanding of the various architectural tactics employed for achieving quality in security analytic systems. |

2.2. Search strategy

According to the guidelines provided in [14] and [15], we defined a search strategy to retrieve as many relevant peer-reviewed papers as possible. Our search strategy is described as follows.

2.2.1. Search method

We used the automatic search strategy to retrieve the potentially relevant papers from six digital libraries: IEEE Xplore, Scopus, ScienceDirect, ACM Digital Library, SpringerLink, and Wiley Online Library. These libraries were searched using the search terms introduced in Section 2.2.2. In order to ensure the identification and selection of as many relevant papers as possible, we complemented the automatic search with snowballing [16, 17].

2.2.2. Search terms

We designed the search string (shown in Fig. 1) according to the guidelines provided in [14]. Our search string consisted of two parts – security and big data processing frameworks. We also incorporated the synonyms and the terms related to the basic two terms. We finalised the search string(s) after ensuring that the pilot searches using the search terms were returning the known papers related to our review. The search terms in the string were matched only with the title, abstract, and keywords of the papers in the digital databases (except SpringerLink which does not allow to restrict the search to specific parts of a paper). We found that several of the retrieved papers were focussed on improving the security of big data and its corresponding technologies (instead of leveraging big data and its corresponding technologies for security that is the focus of this SLR), however, we filtered out those papers during the inclusion and exclusion phase.

> ***TITLE-ABS-KEY*** *((("security" **OR** "intrusion" **OR** "IDS" **OR** "SIEM" **OR** "anomaly detection" **OR** "outlier detection" **OR** "fraud detection" **OR** "network monitoring" **OR** "forensic" **OR** "threat" **OR** "data leakage" **OR** "data breach" **OR** "data theft" **OR** "data exfiltration" **OR** "attack" **OR** "data stealing") **AND** ("hadoop" **OR** "HDFS" **OR** "mapreduce" **OR** "spark" **OR** "flink" **OR** "storm" **OR** "samza" **OR** "mesos")))*

Fig. 1. Search string for this SLR.

2.2.3. Data sources

As previously mentioned, we searched six digital databases, which are shown in Table 2. The IEEE Xplore and ScienceDirect do not support the execution of a query with more than 15 terms. We had to split our query into two parts to make it run on IEEE Xplore and ScienceDirect. Apart from SpringerLink, we ran our query on other databases to match the terms only in title, abstract, and keywords. We have already mentioned that SpringerLink does not support searches in the specific parts of a paper [18]. The limitation of SpringerLink forced us to either restrict our search only to the title of a paper or apply the search string to the whole text of each of the potentially relevant papers. Whilst the former resulted in a very low number of papers, the later returned quite a large number of potentially relevant papers (8483 in total). In order to address this issue, we followed the strategy adopted in [18]. According to this strategy, we examined only the first 1000 papers out of 8483 returned papers. We assert that Scopus is a complement to SpringerLink as it indexes a large number of journals and conference papers in software engineering and computer science [19, 20]. We did not use Google scholar due to its low precision of the results and the tendency of returning a large number of irrelevant papers [21].

Table 2. Database sources.

| Source | URL |
|---|---|
| IEEE Xplore | http://ieeexplore.ieee.org |
| Scopus | https://www.scopus.com/ |
| ScienceDirect | http://www.sciencedirect.com |
| ACM | http://portal.acm.org |
| SpringerLink | https://link.springer.com/ |
| Wiley | http://onlinelibrary.wiley.com/ |

2.3. Inclusion and exclusion criteria

Table 3 shows our inclusion (I) and exclusion (E) criteria that were applied to select the relevant papers from all of the papers retrieved from the digital databases. We focussed on the papers that highlighted the importance of one or more quality attributes and leverage, propose, or evaluate the incorporation of architectural solutions (architectural tactics) for achieving certain quality attributes (e.g., performance, reliability, accuracy, and scalability) in security analytic systems. Due to the design of our search string, it was expected that the search string would return the papers that would have addressed the security of the big data itself or its associated technologies (e.g., security of Hadoop). We excluded such papers during the selection process. We also made sure that any study that leverages big data or its associated technologies for the security of big data itself or its technologies should not be discarded from this SLR.

Table 3. Inclusion and exclusion criteria for this SLR.

| Inclusion criteria |
|---|
| *I1:* A study that is leveraging big data and its corresponding tools and technologies for cybersecurity |
| *I2:* A study that is architecture-based, which means the study should provide architectural solution (i.e., components and architectural model) to the big data security analytics problem |
| *I3:* A study that emphasizes the importance of one or more quality attributes for security analytics |
| *I4:* A study that evaluates the proposed security analytic solution(s) |
| **Exclusion criteria** |
| *E1:* A study that is not peer-reviewed (e.g., position papers, panel discussion, editorials, and keynotes |
| *E2:* A study not written in English |

2.4. Study selection

The papers selected from each digital database at each stage is shown in Fig. 2. The different phases of the selection process are briefly described in the following.

- *Automatic search:* We ran our designed search string on six digital databases and retrieved a total of 4634 papers as a result of running automatic searches.
- *Title-based selection:* We read the title of the papers to quickly decide whether or not each of the retrived papers was relevant to our SLR. In cases, where we were not able to decide about the relevance of a paper only based on its title, the paper was transferred to the next round of selection. Title-based selection reduces the pool of papers from 4634 to 748.
- *Duplication removal:* Scopus indexes papers available in several databases such as IEEE Xplore and ACM, therefore, it was expected that there will be duplicate papers in the pool of 748 selected papers. In this phase, we removed the duplicate papers, which reduced the number of our papers to a total of 516.
- *Abstract-based selection:* We read the abstract of each of the 516 selected papers to ensure that the papers were related to our SLR. Based on the abstract, we discarded 348 papers that brought the pool of our papers to 168.
- *Full-text selection:* The 168 selected papers were passed through full-text selection phase where we read the full text of the papers that fulfilled all the inclusion criteria for this SLR. A total of 69 papers were selected based on reading the full text of the papers.
- *Snowballing:* We applied the snowballing technique [16] to examine the references of the 69 selected papers. Through snowballing, we found 26 potentially relevant papers. We selected 5 papers from the 26 papers based on the inclusion and exclusion criteria. The review included 74 papers.

The reasons for inclusion and exclusion of each of the papers were recorded, which were further discussed among the authors to decide about the ultimate selection of a paper. We did not restrict the selection based on the publication date of a paper. The reason for this was that the incorporation of big data tools and technologies for cybersecurity is a comparatively new research field and as such our search process did not return the papers published before 2009. Appendix A enlists the papers selected for this SLR. We used the terms paper and study interchangeably. Each paper has a unique identifier (S#). For example, the paper "A Cloud Computing Based Network Monitoring and Threat Detection System for Critical Infrastructures" is identified by S10. The system name included in appendix A refers to the name of the security analytic system that was described in a paper. There were systems that were not named by the authors themselves. In case, the system is not named explicitly in the reviewed paper, we followed the strategy adopted in [10] to name the system based on the unique features of the system. We find such naming useful to keep track of different systems when referred or discussed in this SLR.

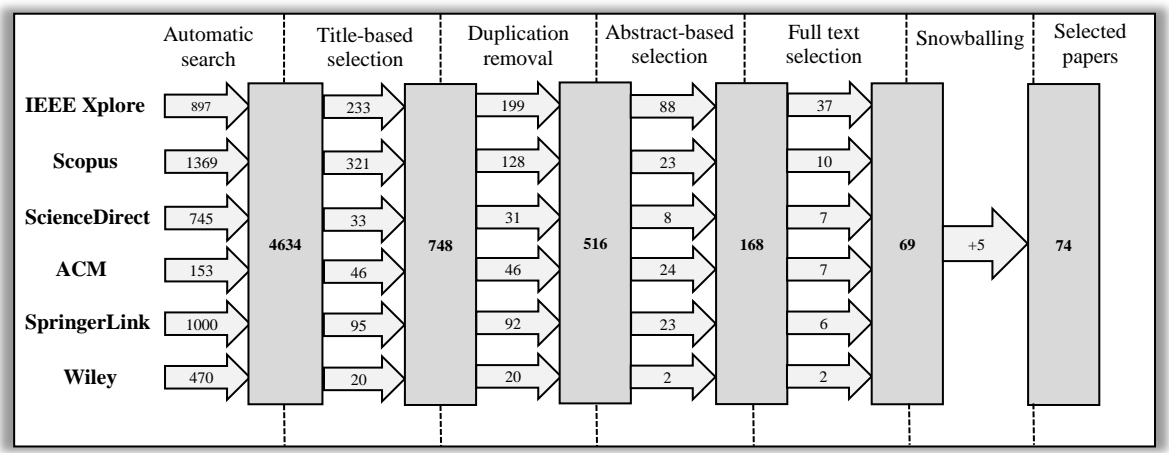

Fig. 2. Phases of study selection process.

2.5. Data extraction and synthesis

This section describes the process of data extraction from the selected papers and the analysis of the extracted data to answer the research questions of this SLR.

2.5.1. Data extraction

We extracted the required data from the selected papers using a pre-designed data extraction form that included the data items envisaged necessary to answer the research questions of this SLR. The data extraction form is shown in Table 4. The data items D5 (publication venue) and D6 (citation count) were included as the indicators of the quality of the selected papers. The extracted data were recorded in an MS Excel spreadsheet for analysis.

Table 4. Data extraction form.

| # | Data item | Description | Research Questions |
|---|---|---|---|
| D1 | Author(s) | The author(s) of the paper | |
| D2 | Year | The publication year of the paper | Demographic data |
| D3 | Title | The title of the paper | |
| D4 | Publication type | The publication type of paper (e.g., conference) | Demographic data |
| D5 | Publication venue | The venue where paper is published | Demographic data |
| D6 | Citation count (google scholar) | The number of times paper is cited according to google scholar | |
| D7 | Application domain | The type of security analytic system (e.g., Intrusion detection system) | Demographic data |
| D8 | Data source(s) | The type of data the system is using for security analytics (e.g., network data) | Demographic data |
| D9 | Big data processing framework | The processing framework system is leveraging (e.g., Hadoop) | Demographic data |
| D10 | Proposed technique | A short summary of the security analysis technique proposed in the paper | |
| D11 | Quality attribute(s) | The quality attribute(s) emphasized in the paper | RQ1 |
| D12 | Rationale for quality attribute(s) | The motivation for highlighting the quality attribute(s) | RQ1 |
| D13 | Architectural tactic(s) | The architectural tactic(s) proposed in the paper | RQ2 |

2.5.2. Data synthesis

We extracted three types of data – demographic data, quality attributes, and architectural tactics. We analyzed the demographic data using descriptive statistics. The results of our analysis of the demographic data have been presented in Section 3. We analyzed the data items D11 (quality attributes), D12 (rationale for quality attributes) and D13 (architectural tactics) using thematic analysis method [22, 23], which is a widely used qualitative data analysis method. We followed the six-step process of thematic analysis for synthesizing the extracted data. The process consists of these steps: (i) *Familiarizing with the data*: The data extracted from the papers and recorded in the excel spreadsheet was well read to get an understanding of the quality attributes critical for security analytics and the architectural tactics for achieving the emphasized quality attributes; (ii) *Generating initial codes:* After gaining the understanding, initial codes were assigned to the rationale for specific quality attributes and the architectural tactics; (iii) *Searching for themes:* The initially generated themes were analyzed to assign specific themes to the rationale identified for each quality attribute and the architectural tactics identified from various papers; (iv) *Reviewing themes:* The themes assigned to the rationale and architectural tactics were compared with each other to decide which themes need to be merged or dropped; (v) *Defining and naming themes:* The names of the themes were reviewed, and themes were renamed wherever required; (vi) *Producing report:* The results of the analysis were reported in the form of quality attributes (Section 4) and architectural tactics (Section 5) for security analytics. It is important to mention that the architectural tactics are not reported in such detail in the primary studies. Therefore, we had to consult other academic sources related to each identified tactic for gathering the data required to report the tactics according to the predefined template reported in Section 5. For example, studies [S30] and [S45] incorporate Data CutOff tactic (Section 5.1.6) and indicate how this tactic can help achieve a good performance. In order to report it according to the designed template, we must consult other academic sources (i.e., [24-26]) to gather the required details about Data CutOff.

## 3. Demographic attributes

This section reports the distribution of the reviewed papers along the years, publication types, big data processing frameworks used, application domains, and data sources leveraged by the reviewed security analytic systems.

3.1. Chronological view

Fig. 3 shows that the reviewed papers were published between 2010 and 2017. Our SLR covers the paper published before *8th May 2017* when the search process for selecting the potentially relevant papers was completed. We could not find many relevant papers before 2013. Finally, the review includes only 6 papers from 2010 to 2012 as compared to 68 papers from 2013 to the middle of 2017. Fig. 3 shows a steady upward growth in the number of papers on security analytic leveraging big data tools and technologies. This is in accordance with the ideas surfaced in the Cloud Security Alliance Workshop [27] where it was anticipated that the role of big data analytics in security will continue increasing. This upward trend can be interpreted from two perspectives. First - the threat landscape is changing from cyber attacks launched by individuals to cyber attacks launched by organized groups supported by influential groups or even states with high budget and sophisticated tools. This change in the threat landscape has motivated the cyber security researchers for exploration to find more advanced ways for detecting cyber attacks. Second - the traditional tools and technologies are unable to deal with the high volume, large size, and heterogeneous nature of security event data; therefore, the adoption of big data tools and technologies is receiving an increasing attention from the cyber security researchers.

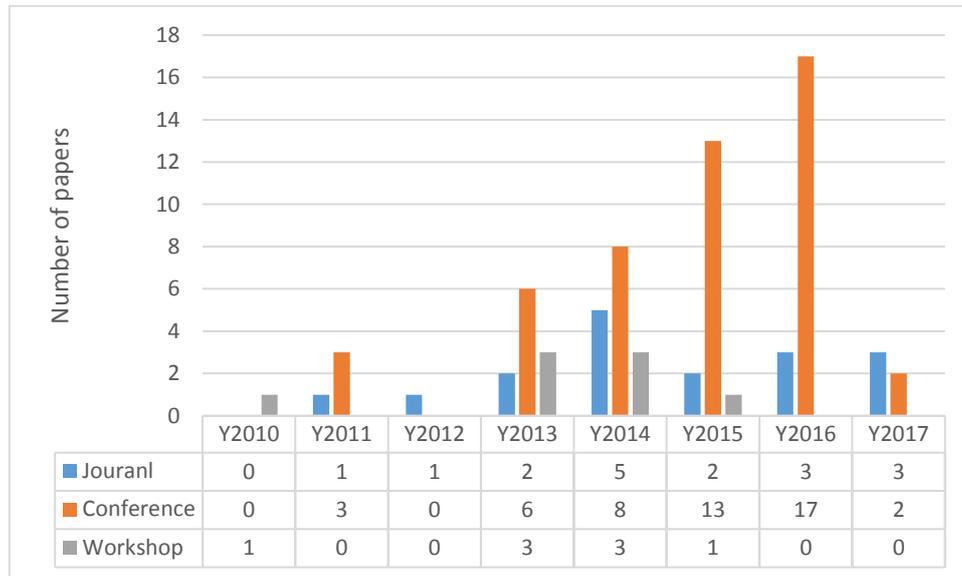

Fig. 3. Number of selected studies published per year and their distribution over types of venues.

3.2. Publication venues and types

Fig 3 indicates that most of the papers (49 papers (66.2%)) have been published in conferences; there were 17 (22.9%) journal papers and 8 (10.8%) workshops papers. The 74 reviewed papers were published in 59 venues, in which BigData Congress, Journal of Supercomputing, and Security and Privacy in Big Data workshop are the leading venues for publishing work on security analytics. These three venues have published three papers each. Apart from these, there are five other venues that have published two papers each. These venues include TrustCom/BigDataSE/ISPA, Conference on Systems, Man, and Cybernetics, Network Operations and Management Symposium, Special Interest Group on Data Communication (SIGCOMM), and Conference on Parallel and Distributed Systems. The rest of the 55 papers were published in 55 different venues. The reviewed papers have been primarily published in three research communities – Big Data (21 paper), Network Communications (17 papers), and Cybersecurity (15 papers). It is interesting to note that only four papers have been published in the Software Engineering related venue. These findings show that researchers with different research backgrounds are interested in security analytics.

3.3. Big data processing frameworks

We have classified the reviewed papers based on the big data processing framework employed in the security analytic systems. The processing framework provides the guidelines for processing the big data. In the reviewed papers, there are three frameworks used – Hadoop[1], Spark[2], and Storm[3]. These frameworks are quite popular as evident from their use by well-known organizations such as Yahoo, Google, IBM, Facebook, and Amazon [28]. A large number of papers report research that use Hadoop (i.e., 49 papers), followed by Spark (i.e. 14 papers), and Storm (i.e., 10 papers) for security analytics. The number and percentage of the papers incorporating different frameworks are shown in Fig. 4a. It was not possible for us to classify one of the papers into either of the three categories based on the information provided in the paper. We also looked into the patterns of these frameworks adopted along the years as shown in Fig. 4b. It is interesting to note that in the first four years (i.e., from 2010 to 2013), only Hadoop has been incorporated whilst for the last four years (i.e., from 2014 to 2017), a steady upward trend can be observed for adopting Spark and Storm. One possible reason for this trend can be the migration from batch processing to stream processing due to the rapidly changing time-sensitive requirements of security analytic systems. The low numbers shown in 2017 is due to the reason that the papers published only before 8th May 2017 are included in 2017.

---

[1] http://hadoop.apache.org/
[2] https://spark.apache.org
[3] http://storm.apache.org/

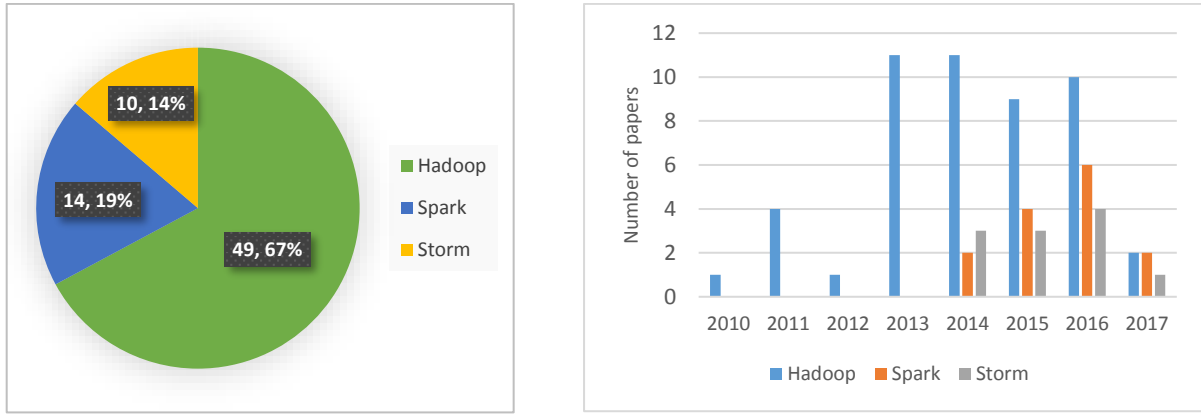

Fig. 4. a) Number and percentage of papers in each category b) Yearly distribution of papers over type of framework.

3.4. Application domains

We have categorized the included papers based on the application domain. For such categorization, we have analyzed the data item D7 in the data extraction form (Table 4). The domain-based categorization is of a potential value for researchers and practitioners who are interested in the domain-specific aspects of security analytics systems. The papers on security analytic systems are primarily of two types – generic and specific. The generic category reports security analytic systems (Intrusion Detection System (IDS) and Alert Correlation) for detecting a variety of attacks. The specific category includes the papers that report security analytic systems designed for detecting a particular type of attack such as Denial of Service (DoS) attack detection. There are several cases where an IDS is evaluated with a particular attack (e.g., [S4] and [S12]), however, we have included such systems in the IDS category only. As shown in Table 5 that the 74 reviewed papers have been categorized into nine groups. The majority of the reviewed papers belong to IDS category (i.e., 34 papers) followed by unclear (11 papers) and alert correlation (i.e., 8 papers). The unclear category includes the papers that cannot be explicitly (i.e., without authors' interpretation) categorized into either of the remaining 8 groups.

Table 5. Distribution of application domains of the reviewed papers

| S. No | Application domain | # of papers | Papers |
|---|---|---|---|
| 1 | Intrusion Detection System | 34 | S1, S3, S4, S5, S8, S9, S12, S14, S15, S18, S21, S26, S27, S29, S31, S32, S34, S35, S41, S44, S46, S48, S51, S54, S55, S57, S59, S63, S64, S66, S67, S72, S73 |
| 2 | Alert Correlation | 8 | S6, S7, S11, S16, S19, S38, S39, S65 |
| 3 | DoS Detection | 7 | S2, S20, S28, S33, S43, S61, S62 |
| 4 | Botnet Detection | 4 | S22, S23, S36, S69 |
| 5 | Forensic Analysis | 2 | S30, S53 |
| 6 | APT Detection | 2 | S37, S47 |
| 7 | Malware Detection | 4 | S43, S45, S60, S68 |
| 8 | Phishing Detection | 2 | S17, S50 |
| 9 | Unclear | 11 | S10, S13, S25, S40, S49, S52, S56, S58, S70, S71, S74 |

3.5. Data sources

A security analytic system can collect data from a number of data sources. Based on the data source, security analytic systems can be classified into three categories – (1) *Network,* (2) *Host,* and (3) *Hybrid*. The number and the percentages of the papers pertaining to each category are shown in Fig. 5. The systems belonging to the first category (i.e., network) collect data from network infrastructure of an organization. Within the network infrastructure, a security analytic system can leverage different kinds of data such as Netflow data, packet data, honeypot data, IDS log data, and firewall logs [29]. The systems belonging to the second category (i.e., host) collect data from an organization's host machines. These data sources include but not limited to operating system logs, system call logs, web server logs, email logs, and windows event logs. The systems belonging to the third category (i.e., hybrid) collect data from both network and host machines. Fig. 5 shows that most of the systems (i.e., 55) in our review rely on the data collected from the network, followed by hybrid (i.e., 13), and only 6 systems leverage host data for security analytics. One possible reason for such a focus on the network-based security analytic systems could be that unlike host-based systems that protect only a specific host within an organization, the network-based systems take into consideration the security of an entire organization's network. The pros and cons of network-based and host-based security analytic systems have already been well explored. Interested readers can refer to [30-32].

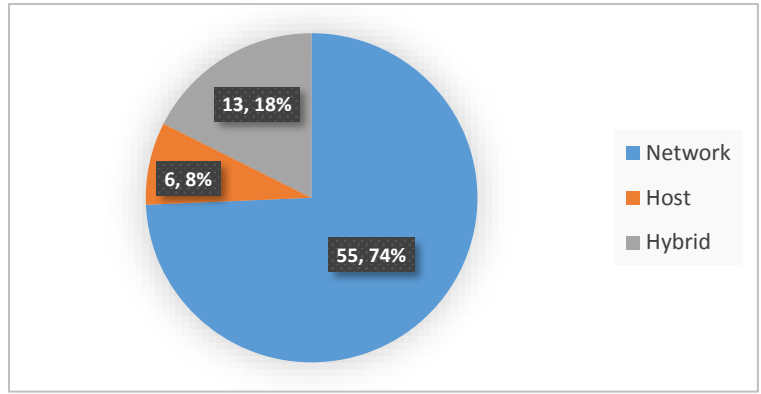

Fig. 5. Number and percentage of papers distributed over type of data source.

## 4. RQ1: Quality attributes

This section reports the results based on analysis of the data about critical quality attributes reported for security analytics and their respective rationale. The analysis is meant to answer RQ1, "*Which are the most important quality attributes for security analytic systems?*". We answer this question from two perspectives: (1) We present the statistical analysis of the quality attributes identified in the reviewed papers to understand the depth of emphasis on various quality attributes and (2) We report the identified motivation to justify that why these particular quality attributes have been highly emphasized for security analytic systems.

4.1. Statistics of the quality attributes

This section reports the findings from the analysis of item D11 of data extraction form (Table 4), which records the quality attributes emphasized in the reviewed papers. Table 6 presents the number, percentage, and identifiers of studies emphasizing particular quality attributes. These are the quality attributes that are emphasized, highlighted, or considered important for security analytic systems. Considering a quality attribute important does not mean that the paper is focussed on achieving the particular quality attribute as well. For example, [S15] highlights interoperability as a critical quality attribute for security analytic systems, however, it does not devise any strategy or mechanism for achieving interoperability. It is worth noting that a single paper might be emphasizing multiple quality attributes. For example, [S2] considers performance, accuracy, scalability, and reliability critical quality attributes for a security analytic system. It can be seen in Table 6 that performance, accuracy, scalability, and reliability are the most important quality attributes with a total number of 67, 43, 40, and 23 papers respectively. It is important to mention that apart from the reported 12 quality attributes, our reviewed papers also highlighted some other quality attributes, however, such quality attributes are found in less than three studies, therefore, those quality attributes are not explicitly shown in Table 6. These quality attributes include design simplicity [S6], algorithm expressiveness [S6], modularity [S10], uniform programmability [S17], and reusability [S74].

4.2. Quality attributes definition for security analytic systems

This section reports our analysis of the data item D12. The objective is to clarify the definition, use, and emphasis on the 12 quality attributes presented in Table 6. Such clarification will help us to answer the question that why these particular quality attributes are more important for a security analytic system. For each quality attribute, we provide a definition and motivation for their achievement in a security analytic system.

Table 6. Papers emphasizing quality attributes

| Quality attribute | # of papers | % of papers | Paper identifiers |
|---|---|---|---|
| Performance | 67 | 90.5 | S1, S2, S3, S4, S5, S6, S7, S8, S9, S10, S11, S12, S13, S14, S15, S16, S17, S18, S19, S20, S21, S23, S24, S26, S27, S29, S30, S31, S32, S33, S34, S35, S36, S37, S38, S39, S40, S41, S42, S43, S44, S45, S46, S47, S48, S49, S50, S51, S52, S53, S54, S55, S56, S57, S58, S60, S61, S62, S63, S64, S65, S66, S67, S70, S71, S73, S74 |
| Accuracy | 43 | 58.1 | S1, S2, S5, S8, S9, S10, S12, S14, S15, S18, S20, S21, S22, S23, S24, S25, S26, S27, S28, S29, S30, S31, S36, S37, S40, S41, S42, S43, S46, S47, S49, S55, S57, S58, S59, S60, S61, S62, S64, S66, S67, S70, S73 |
| Scalability | 40 | 54.0 | S2, S3, S5, S6, S7, S8, S9, S10, S12, S15, S16, S17, S18, S19, S21, S22, S24, S25, S29, S30, S31, S38, S42, S45, S47, S48, S49, S51, S52, S53, S54, S56, S57, S61, S65, S66, S68, S70, S73, S74 |
| Reliability | 23 | 31.0 | S2, S6, S8, S10, S13, S17, S18, S19, S20, S30, S31, S43, S48, S49, S52, S54, S55, S59, S61, S62, S65, S66, S71 |
| Usability | 18 | 24.3 | S10, S17, S19, S24, S29, S30, S40, S41, S51, S52, S53, S56, S65, S66, S68, S72, S74 |
| Interoperability | 15 | 20.2 | S10, S15, S16, S17, S19, S35, S39, S40, S42, S46, S47, S50, S51, S54, S72 |
| Adaptability | 11 | 14.8 | S1, S7, S10, S15, S20, S28, S28, S39, S42, S61, S74 |
| Modifiability | 7 | 9.4 | S7, S10, S12, S29, S40, S44, S56 |
| Generality | 7 | 9.4 | S7, S10, S12, S19, S22, S29, S51 |
| Privacy assurance | 7 | 9.4 | S15, S19, S25, S30, S36, S42, S49 |
| Security | 4 | 5.4 | S10, S49, S53, S73 |
| Stealthiness | 3 | 4.05 | S9, S21, S51 |

***Performance*** is a measure of how quickly a system responds to user inputs or other events [33]. Performance encompasses several aspects of a system such as response time, throughput, and latency. In case of security analytics, response time is a highly critical aspect as highlighted in [S1], [S17], and [S20]. These systems are expected to respond in real-time to cyber attacks in order to thwart the attacks or take mitigation actions after an attack. However, it is quite challenging to achieve a real-time response in security analytic systems due to several reasons. The volume of security event data that these systems have to analyze is quite massive [S1], [S7], [S9], [S26] and [S48]. Similarly, due to widespread adoption of Internet, the security event data is generated at a very high speed as mentioned in [S3] and [S46]. The quality of the collected security event data is another hurdle in the face of real-time security analytics. The collected data needs to be intensively pre-processed. For example, the data dimensionality needs to be reduced [S44] and data should be converted from binary format to text format [S16]. Furthermore, the current threat landscape requires a widespread correlation and contextualization of security event data to detect sophisticated attacks designed to execute over a period of time. In order to deal with such threats, a highly complex computation is incorporated, which makes the real-time analytics a challenging task [S19], [S26].

***Accuracy*** is a measure to which a system provides the right results with the needed degree of precision [34]. To make it specific to security analytics, it is the ratio of the total number of correctly detected attacks to the total number of attacks [S31]. As evident from Table 6, accuracy is a highly critical quality for security analytic systems. The consequences of letting an attack go undetected can be catastrophic. Some recent examples of such catastrophes can be found in [35] where successful execution of cyber attacks caused almost catastrophic damages to systems in the governmental and private organizations. In the context of security analytics, accuracy becomes even more critical as a security analytic system is expected to only detect and resist attacks and does not become a shield in the face of any legitimate access request (i.e., false positives) [S1], [S46]. A security analytic system generating a large number of false positives may not be deployed in an organization.

***Scalability*** is a measure of how easily a system can grow to handle more user requests, transactions, servers, or other extensions [33]. Table 6 reveals that around 54% of the papers in our review consider scalability as an important quality for security analytic systems. It is quite challenging to predict the volume and velocity of data to be processed by a security analytic system even if we know the dynamics of an organization [36][S7], therefore, such a system need to be highly scalable to deal with any types of irregularities. Furthermore, it is well understood that sophisticated cyber attacks (e.g., APTs) are executed over a period of time, which can push a system to consider the analysis of a constantly growing size of security event data [S17], [S31], [S74]. A well-known example of such an attack is Operation Shady RAT [37], which started in 2006 and continued until discovered in 2011. The attack affected around 72 different organization around the world including US government and United Nations. A system designed for the analysis of a limited data size will not be able to detect such advanced attacks.

The advancement in cyber attacks demands bringing in more and more sources (e.g., database access and users' activities) within an organization under the umbrella of monitoring and observation [S56]. Such an increase in the monitoring capability can only be achieved if the system is capable to scale with the increase in demand. It is quite apparent from the mentioned facts that a security analytic system should be highly scalable, however, achieving scalability in security analytics is quite challenging as highlighted in the papers reviewed for this study. Some prominent hurdles in the face of scalability include: (a) inefficient communication among a large number of processors in a distributed setup [S57]; (b) unfair load-balance among the computing nodes [S57] and (c) the choice of centralized data storage [S50].

*Reliability* is a measure of how long a system runs before experiencing a failure [33]. Reliability quality attribute is fourth on the list in terms of importance in the reviewed papers (Table 6). A security analytic system might be vulnerable to a number of failures. A compute node experiencing heavy processing load can crash due to the large size of the security event data directly feeding into a node [S1], [S18]. In addition to reliable data processing, reliable data collection is also quite crucial for security analytics [S33]. By reliable data collection, we mean that the data collector should cope with the speed of data so that all security event data (e.g., NetFlow data [38]) can be captured. For example, it is quite possible that a single NetFlow may contain information significant for detecting an attack and letting such a NetFlow go uncaptured and unprocessed means letting the attack go undetected. It is worth mentioning that the traditional data collectors (e.g., Wireshark) are lagging behind in efficiently collecting data under peak conditions (e.g., DoS attack) [S63]. Furthermore, the inherent design limitations of software and hardware also need some attention while designing a security analytic system as it may lead to several types of malfunctioning [S71].

*Usability* is a measure of how easy it is for people to learn, remember, and use a system [33]. Considering the spontaneous nature of cyber attacks where delaying the response by a few seconds can be consequential, it is important that a security analytic system is user-friendly [S10]. A security analytic system should facilitate security experts or administrators by visualizing the results (i.e., threat alerts) in a human-friendly way so that the required mitigation action can be taken [S17], [S29], [S41], [S58], [S66]. In addition to facilitating the response to an attack, a system should visualize the status of the system to the administrator in a convenient way, hence, enabling the administrator to make well-informed decisions about data storage, transmission, and analysis [39]. Similarly, a security analytic system might generate hundreds of alerts in a short period of time and a security administrator might not be able to handle all of them. Therefore, the system should incorporate some mechanism that enables a security administrator to focus on the most dangerous alerts [S41].

*Interoperability* is a measure of how easily a system can interconnect and exchange data with other systems or components [33]. According to Table 6, around 20.2% of the reviewed studies consider interoperability as an important quality attribute for security analytic systems. Unlike other data-intensive systems, the interoperability requirement is more critical in the domain of security analytics from several aspects. A security analytic system needs to integrate with other security systems (e.g., firewall or antivirus) to enhance the overall security spectrum of an organization [S14], [S17]. A collaborative Intrusion Detection System (IDS) is a well-established example where multiple IDSs collaborate with each other for detecting sophisticated attacks [40]. The security analytic systems connect to a variety of sources (e.g., network devices and host machines) for data collection for which limitation on interoperability can be problematic. Motivated by its data-intensive nature, a security analytic system need to interoperate with a variety of databases such as Oracle, MySQL, MsSQL [S10], [S16], [S47]. Moreover, a security analytic system should also be able to integrate with specialized hardware (e.g., GPGPU) for achieving fast data processing capability [S35].

*Adaptability* is the measure of how easily a system adapts itself to different specified environments using only its own functionality [34]. It is expected that a security analytic system automatically adjusts itself to various kinds of changes as apparent from our finding shown in Table 6. A security analytic system incorporating multiple data sources deals with different data formats (such as formatted text and binary data) so the system should be able to automatically adjust itself with the data format without affecting the overall performance of a system [S1], [S39], [S74]. Moreover, a system should automatically change the security policies according to the requirements [S10]. For example, more in-depth monitoring of insiders during working hours as compared to non-working hours. Similarly, the network of an organization frequently experiences changes (e.g., change in network topology), therefore, a security analytic system should adjust itself to the changes in a network environment [S15].

*Modifiability* is a measure of how easy it is to maintain, change, enhance, and restructure a system [33]. It is worth mentioning that modifiability relates to manual modification by a user while adaptability is the automatic adjustment by a system itself without any involvement of a user. Several changes are beyond the control of a system itself and require an input from a user, therefore, around 9.4% of the reviewed papers consider that modifiability is a critical quality. In the context of security analytics, a number of different modifications are required from time to time. A signature-based security analytic system needs to be updated with the latest and emerging attack patterns [S7]. Similarly, with the advancement of algorithms and computational models, a system needs to be modified to employ the advanced algorithms and models as highlighted in [S10], [S12], [S56]. Furthermore, with the flourishment of big data tools and technologies, the security analytic systems should be flexible enough to easily incorporate the upcoming tools and technologies [S17], [S40]. Modifiability will be a key attribute to enable a security analytic system to advance itself at the same pace with which the current threat landscape is advancing.

*Generality* is a measure of the range of attacks covered by a security analytic system [S7]. As evident from Table 5 that different security analytic systems target the detection and prevention of different types of attacks. It is also mentioned in the application domains (i.e., Section 3.4) that some of the security analytic systems (e.g., IDS, alert or correlators) are generic and cover a wide variety of attacks. Security analytic systems that are designed to detect specific attacks are more accurate as compared to systems that try to detect a variety of attacks [41]. However, the systems specific to a particular attack are less flexible to accommodate the changing nature of attacks. Several of the reviewed papers ([S7], [S12], [S19], [S22]) have revealed that a security analytic system has to be generic enough to counter the variety, complexity, and sophistication of different types of cyber attacks. The trade-off between generality and specificity should be established based on the specific security requirements of an organization. If the security experts believe that their organization is vulnerable only to a particular type of attack (e.g., botnet attack), then a specific security solution should be preferred over a generic solution.

*Privacy assurance* is the measure of the ability of a system to carry out its business according to defined privacy policies to help users trust the system [42]. Several of the reviewed studies ([S15], [S19], [S25], [S30], [S36], [S42], [S49]) highlight the importance of privacy assurance for security analytics. In the context of security analytics, privacy assurance is quite critical because a number of security analytic systems capture, store, and process packet payload, which contains personal data of users. For example, one of the reviewed papers i.e., [S17] employs a technique called content inspection [43], which in addition to the header information, also capture, store, and process payload of network packets. According to the European data laws and regulations [44], it is not allowed to store and process the packet payload without the relevant users' consent. Therefore, any security analytic system intending to analyze the personal data of users should undergo a privacy agreement with its users.

*Security* is the measure of how well a system protects itself and its data from unauthorized access [33]. Considering the fact that the goal of a security analytic system is to protect an organization's overall infrastructure, such a system itself also needs to be secure. For instance, if an attacker can get access to a security analytic system; he/she may modify the monitoring rules. In another scenario, an attacker modifies some critical bits in the data during its transmission between two services (e.g., data collection and processing) of a security analytic system. Such modifications achieved via unauthorized access will disable a security system to accurately detects attacks. This issue is highlighted by some papers (i.e., [S10], [S49], [S53], [S73]) in our review as shown in Table 6.

*Stealthiness* is the measure of the ability of a security analytic system to function without being detected by an attacker [41]. If an attacker knew the whereabouts of a security system, it is very likely that the first thing, an attacker will do after getting into the organization's infrastructure, is to disable it's security system [45]. Therefore, it is important for a security analytic system to operate in a stealthy mode, whereby an attacker does not even know that a security system is in place for monitoring the attacker's activities. In our review, only three studies (i.e., [S9], [S21], [S51]) emphasize the significance of stealthiness for a security analytic system.

## 5. Architectural tactics

This section reports the results of analysis of data on architectural tactics for security analytics systems. This part of the analysis is meant to answer RQ2, "*What are the architectural tactics for addressing quality concerns in security analytic systems?*". An architectural tactic is a design strategy that influences the achievement of a quality attribute [9]. The architectural tactics presented in this section have been elicited from the reviewed studies based on (i) the quality attribute(s) explicitly stated; (ii) the quality attribute(s) inferred from the reported architecture for the proposed system and (iii) the components and their relationship determined in the architecture of the proposed system. Fig. 6 shows the identified tactics. We have identified six tactics for performance, four for accuracy, two for scalability, three for reliability, and one for security and usability each. These tactics are reported using the following template.

- *Introduction:* brief explanation of how tactic achieves the desired quality attribute;
- *Motivation:* rationale behind why the tactic needs to be incorporated in an architecture design;
- *Description:* detailed explanation on how the various components of a system interact to achieve the desired quality attribute and the architecture diagram highlighting the components related to a tactic;
- *Constraints:* necessary conditions for incorporating the tactic in the existing architecture of a system;
- *Example:* one or more systems from the reviewed studies demonstrating the application of the tactic;
- *Dependencies:* whether or not a tactic depends upon another tactic(s) for its incorporation in the system;
- *Variation (optional):* slightly modified form of the original tactic

It is worth mentioning that we have tried to follow the same diagrammatic style for reporting all architectural tactics. Because of this, we have to modify the diagrams in a way that can best explain the specific tactic instead of explaining the entire architecture of a security analytic system. For example, the component responsible for feature selection and extraction is not explicitly shown in tactics other than the Feature Selection and Extraction tactic (Section 5.1.3); it does not mean that this component does not necessarily exist in the diagrams for other tactics. Similarly, the rationale behind showing the data collection as a distributed process (in Data Ingestion Monitoring tactic (Section 5.4.1) and Secure Data Transmission tactic (Section 5.5.1)) and as a centralized process is to suit the tactic and explain its key points.

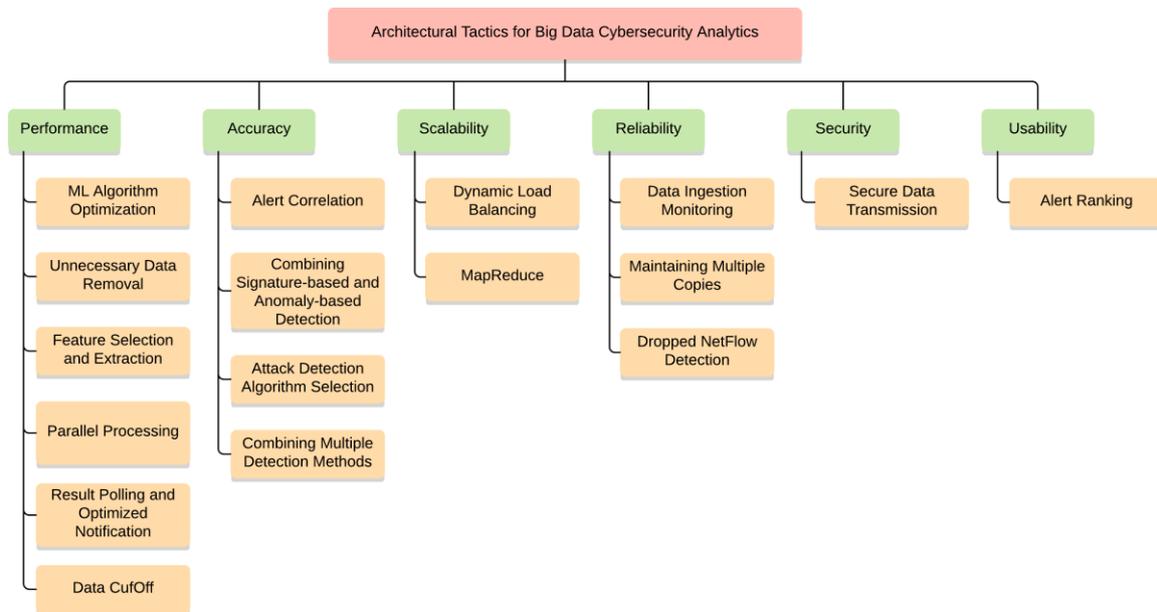

Fig. 6. Architectural tactics for Security Analytics Systems classified based on the relevant quality attributes

5.1. Performance

This section reports the architectural tactics related to Performance quality attribute.

**5.1.1. ML algorithm optimization**

*Introduction.* The ML Algorithm Optimization tactic is found in all the papers as all of the security analytic systems leverage some type of Machine Learning (ML) algorithm for analyzing the security event data. The objective of this tactic is to highlight the role of algorithms in improving the performance of a system and provide some guidelines for selecting the algorithm that is most efficient in terms of computational complexity.

*Motivation.* The two most important factors related to the performance of a security analytic system are the type of input data and the employed ML algorithm [53]. A number of ML algorithms are available that can be leveraged in a security analytic system. These ML algorithms range from supervised learning (e.g., Logistic Regression, Support Vector Machine, Naïve Bayes, Random Forest, and Decision Trees) to unsupervised learning algorithms (e.g., K-means and Neural Networks). For details on the ML algorithms used in the security analytics systems, readers should refer to [53]. When selecting a ML algorithm, several factors need to be considered. These factors include the time complexity, incremental update capability, offline/online mode, and generalization capacity of the algorithm, and most importantly the impact of the algorithm on the detection rate (accuracy) of a system. Due to the diverse role of the algorithm, it is quite challenging to pick the most appropriate and efficient algorithm.

*Description.* The main components of a ML Algorithm Optimization tactic are shown in Fig. 7[4]. The *data collection* component collects security event data for training a security analytic system. The training data can be collected from sources within the enterprise where a system is supposed to be deployed as depicted in Fig. 7 or an already available dataset such as KDDcup99 can be used as training dataset. After collecting the training data, the *data preparation* component prepares the data for training the model by applying various filters. Next, the selected *ML algorithm* is applied to the prepared training data to train an attack detection model. The time taken by the algorithm to train a model (i.e., training time) varies from algorithm to algorithm. Once the model is trained, it is tested to investigate whether the model can detect cyber attacks. For testing the model, data is collected from the enterprise as shown in step-4. The testing data is filtered through the *data preparation* component and fed into the *attack detection model*, which analyzes the data for detecting attacks based on the rules learned during the training phase. The time taken by an *attack detection model* to decide whether a particular stream of data pertains to an attack (i.e., decision time) depends upon the employed algorithm. The result of the data analysis is displayed to the user through *visualization* component.

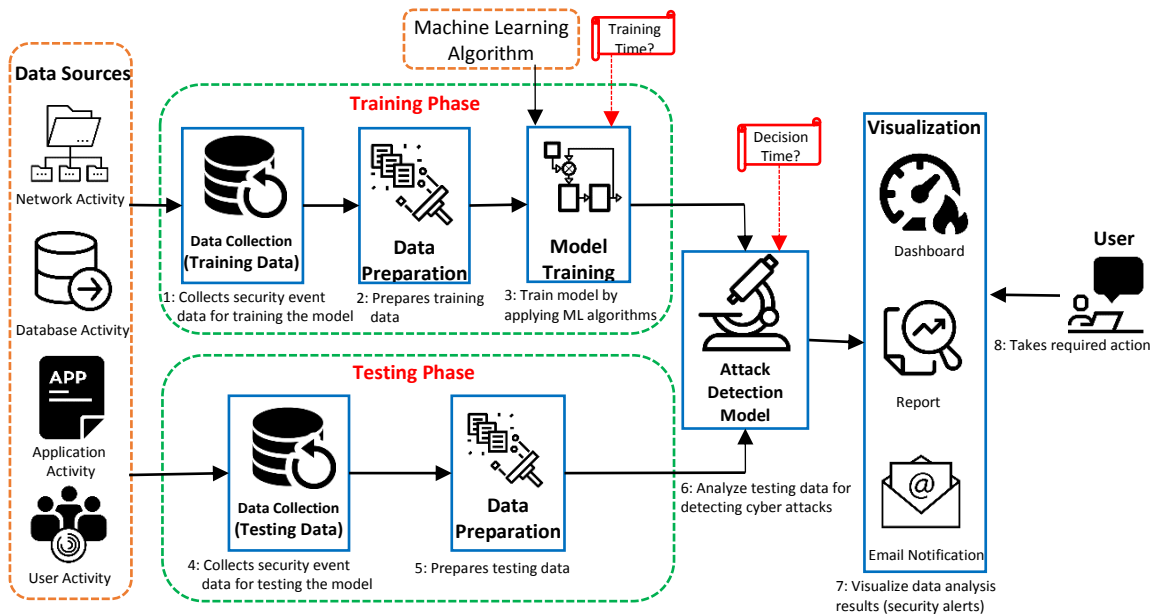

Fig. 7. Algorithm optimization tactic

*Constraint.* A software architect needs to keep an eye on several things during the selection and incorporation of an optimized algorithm.

- The empirical comparison for ML algorithms in [54] suggests that performance of ML algorithms is not consistent in different problem domains. Therefore, an ML algorithm may perform well for one type of security analytics (e.g., detecting DoS attack) but may not perform well in another type of security analytics (e.g., detecting brute force attack).
- The selection of an algorithm is challenging in the sense that in addition to performance, it affects other system's qualities such as accuracy, complexity, and understandability of the final result. For example, Cheng et al. [55] compare SVM with Extreme Learning Machine (ELM) in terms of accuracy and performance. It is found out that SVM generates more accurate results but is computationally expensive. On the other hand, ELM generates less accurate results but is more light-weight. A reasonable trade-off should be established among various system's qualities while selecting the algorithm.

---
[4] Source of icons in the diagrams: https://thenounproject.com/

- The selection of algorithm also depends upon the working mode (online or offline) of a security analytic system. Generally, algorithms having a time complexity less than $O(n^3)$ are considered acceptable for online mode while algorithms with a complexity of $O(n^3)$ and above are slower and suits only offline analysis mode [53].

*Example.* As mentioned all of our included systems leverage ML algorithms, however, here, we report the findings from a couple of papers to exhibit the role of an optimized algorithm in improving the performance of the security analytic system.

- Spark-based IDS Framework [S9]: This security analytic system compares the training time and decision time of five ML algorithms namely Logistic regression, Support vector machine, Random forest, Gradient boosted decision trees, and Naïve Bayes. With KDDCup99 dataset, Naïve Bayes shows the best training time (i.e., 79.5 sec) and SVM shows the worst training time (i.e., 479.12 sec). On the other hand, with respect to decision time, SVM shows the best decision time (i.e., 10 sec) and Gradient boosted decision tree shows the worst decision time (i.e., 22.2 sec).
- Ultra-High-Speed IDS [S14]: Here, the security analytic system is tested with six ML algorithms to investigate the training time and decision time of each algorithm. The algorithms are Naïve Bayes, SVM, Conjunctive rule, Random Forest, J48, and RepTree. It is found that both in terms of training time and decision time RepTree is the most efficient followed by J48.
- Cloud-based Threat Detector [S10]: The system has been implemented with two ML algorithms – k-mean and Naïve Bayes, to explore the training time taken by both the algorithms to train a system. It is observed that with 500 GB training data, k-means takes around 60 secs while Naïve Bayes takes around 92 secs to train a model.

*Dependencies.* The ML Algorithm Optimization tactic requires Unnecessary Data Removal tactic (Section 5.1.2) and Feature Selection and Extraction tactic (Section 5.1.3) to help bring collected data into a refined form. After the application of these tactics, the ML Algorithm Optimization tactic can be efficiently applied to the refined data to quickly train a system and detect attacks. The ML Algorithm Optimization tactic needs to be incorporated alongside Attack Detection Algorithm Selection tactic (Section 5.2.3) as these two tactics establish the trade-off between the effects of the ML algorithm on performance and accuracy.

### 5.1.2. Unnecessary data removal

*Introduction.* The Unnecessary Data Removal tactic have been found in four of the reviewed studies ([S52], [S58], [S66], [S67]). This tactic removes the unnecessary data from the dataset of security event data that is supposed to be processed by the data analysis module of a system to detect cyber attacks. The subset of security event data that does not contribute to the detection process is termed as unnecessary data. The removal of such data from the dataset reduces the size of the dataset, which decreases the processing time.

*Motivation.* Security-critical data is collected from a variety of sources within an enterprise to detect cyber attacks. However, not all of the collected data contributes to the detection process. For example, a network sniffer captures zero-byte data that is not useful in detecting cyber attacks as zero-byte flows are primarily used for the handshaking in a TCP/IP connection [46]. Therefore, such useless data should be removed from the rest of the data captured by the network sniffer.

*Description.* Fig. 8 shows the various components of the Unnecessary Data Removal tactic. *Data collector* component collects security event data from various sources within an enterprise. The *data storage* component stores the collected data. Before forwarding data for analysis, data is intercepted by the *data cleaning* component, which filters the data and removes unnecessary data from the dataset. After the unnecessary data is removed, the rest of the data is forwarded to the *data analysis* component that analyzes the data to detect cyber attacks. Finally, the analysis results are visualized through the *visualization* component.

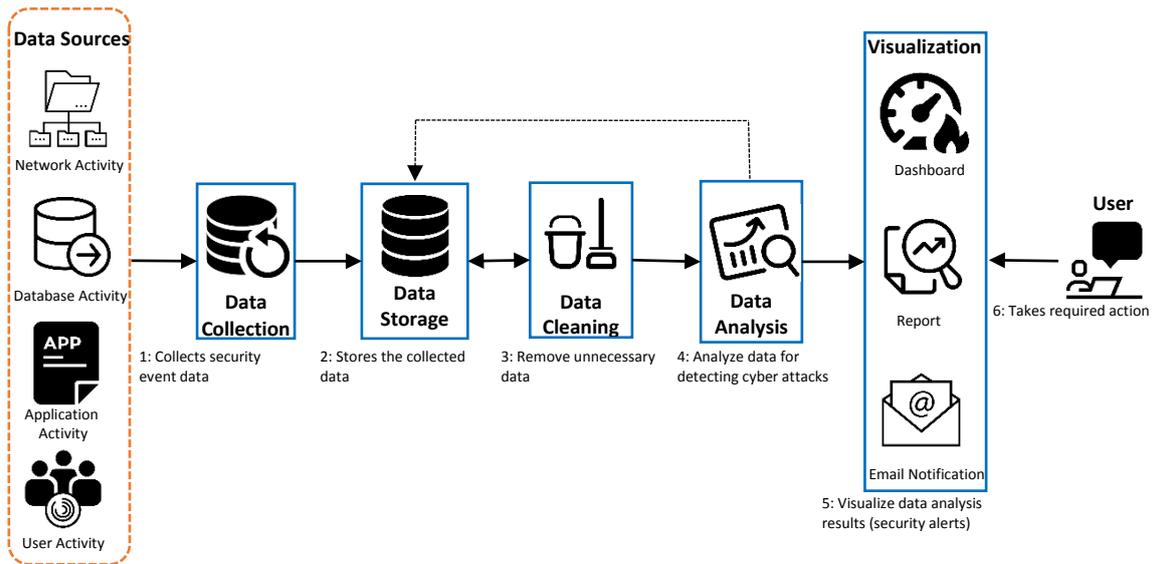

Fig. 8. Unnecessary data removal tactic

*Constraints.* This tactic requires that the data cleaning functionality is not computationally expensive, otherwise, it would cancel the benefit of quick response acquired through reduction of data size. In addition, the rules for filtering out unnecessary data need to be designed carefully to ensure that data that is critical for accurately detecting cyber attacks does not get filtered out. These data filtration rules vary from situation to situation as data that is of significant value in one situation may not be significant in another situation [S67].

*Example.* The following two systems incorporate Unnecessary Data Removal tactic.

- Dynamic Time Threshold [S58] analyzes security event data collected from various sources that include web access log, network log, host log, and network behaviour log. This system incorporates Unnecessary Data Removal tactic to remove log record with URL suffix JPG, JPEG, GIF, WAV, CSS and JS. These log records automatically get stored in the web access log when a user requests a web page associated with CSS, JS, or image data. However, these log records do not contribute to the threat analysis process, therefore, these are considered unnecessary and are removed.
- Batch and Stream Analyzer [S66] collects and analyze NetFlow data for detecting cyber attacks. This system employs Unnecessary Data Removal tactic for removal of zero-byte flows captured in the NetFlow data. As mentioned, zero-byte flows are used for handshaking in TCP/IP connection and have no relevance to threat detection, therefore, such data is removed from the rest of the Netflow data.

*Dependencies.* The Unnecessary Data Removal tactic requires Parallel Processing tactic (Section 5.1.4) to speed up the process of removing unnecessary data from the raw data collected from different sources.

*Variations: Removal of duplicates.* This tactic as described removes data that does not contribute to attack detection process. However, collected data also contains lots of duplicate records (i.e., records pertaining to same activity in a certain period of time). Including a single instance of such data is valuable for attack detection but analysing redundant records puts an extra burden on the computational process without any valuable contribution to the detection process. Therefore, specific representative instances of such records should be included, and duplicates should be removed before forwarding data to the analysis module. The removal of duplicates not only supports fast processing but also gives a much clearer view of the activities directed towards or within a network [S6]. Compression Model for IDS [S21] implements this tactic on training data for IDS. The system uses affinity propagation [47] to remove duplicates from the training data and so extract a small dataset from a large-scale dataset. It was found that the incorporation of removal of duplicate tactics along with horizontal compression enhances the efficiency by 184 times with less than 1% negative effect on the attack detection accuracy. Multistage Alert Correlator [S6] incorporates this tactic for removing duplicate alerts in an alert correlation system designed to correlate individual alerts to determine attack patterns that can be used for predicting future attacks.

### 5.1.3. Feature selection and extraction tactic

*Introduction.* The Feature Selection and Extraction tactic has been found in [S9], [S10], [S12], [S13], [S14], [S15], [S31], [S32], [S33], [S55], [S57]. This tactic selects and extracts the most relevant features from the network traffic data that can be analyzed for detecting cyber attacks. The incorporation of this tactic helps to reduce storage volume, increase data processing speed, and reduce the complexity of the data. This tactic not only improves the performance of a system but also contributes to improving the detection accuracy.

*Motivation.* According to a Cisco report, global IP traffic was 1.2 Zettabytes in 2016, which is expected to reach 3.3 Zettabytes in 2021 [48]. Each IP traffic record contains more than 40 features (such as source IP address, destination IP address, source port, and destination port). For example, the traffic records contained in Knowledge Discovery in Databases cup 1999 (KDDcup99) dataset [49] consist of around 41 features while traffic records in Centre for Applied Internet Data Analysis (CAIDA) dataset [50] has around 50 features. Analysing such a large size of network traffic data using all features is an inefficient approach from two perspectives. First, it requires very high computational power to analyze such large size data even in offline analysis. Many enterprises with limited resources cannot afford to allocate such high computational capability to security analytics systems. Second, there are several features among the captured features for each record that do not contribute to the attack detection process, rather reduces the detection accuracy of the security analytic system. Therefore, it is imperative to carefully select and extract specific features from the captured network traffic for security analytics.

*Description.* The major components of the Feature Selection and Extraction tactic are shown in Fig. 9. The *data collection* component collects network traffic data from different sources (e.g., router, switch, or firewall). The data captured by data collection module is uncleaned and each record contains more than 40 features. The *feature selection* module selects specific features from a large number of features for each IP traffic record. There exist several techniques for selection of features from records. The choice and number of features also vary from one case to another. After selecting the features, the specific features are extracted from the records by the *feature extraction* module. The *feature selection* and *feature extraction* modules leverage parallel processing capabilities to speed up the selection and extraction process. The extracted feature dataset is passed onto the *data analysis* module that analyzes the reduced dataset to detect cyber attacks. In case of an attack situation, alerts are generated that can be viewed by the user (e.g., security expert or network administrator) through *visualization* component. Once such attack alerts come under notice, a user or enterprise can take necessary steps to prevent or mitigate the effects of the attack.

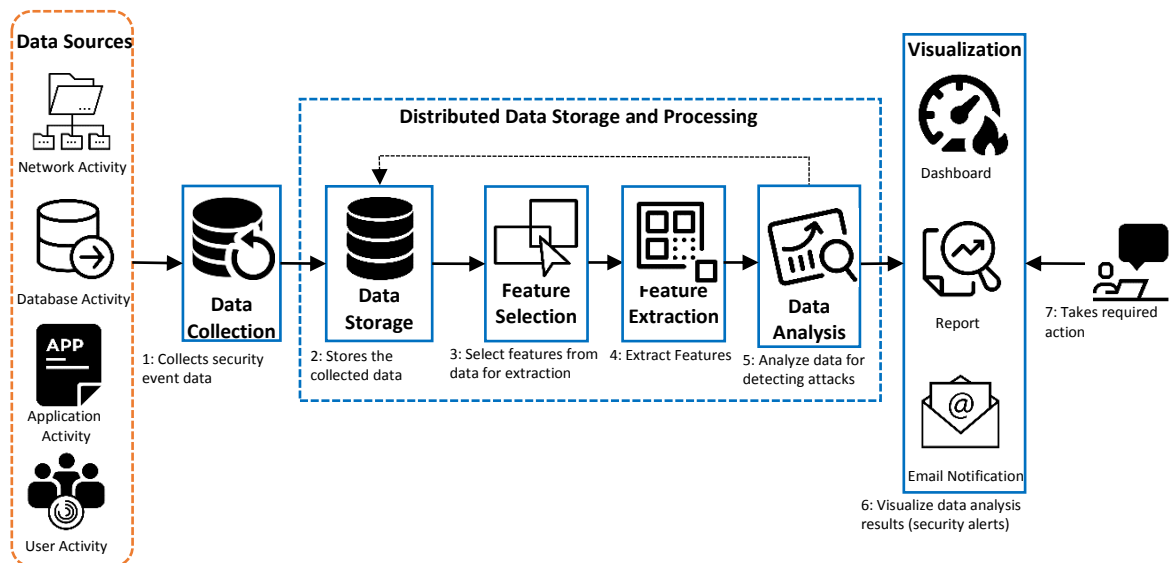

Fig. 9. Feature selection and extraction tactic

*Constraints.* The Feature Selection and Extraction tactic facilitates real-time security analytics, however, if critical features are missed in the selection process, it may severely affect the detection accuracy of a system. There must be a careful approach to ensure that any feature that contributes to the detection process should not be missed. This condition becomes more relevant in cases where the network traffic is constantly evolving due to emerging technologies and new attacks. In addition, the feature selection and extraction operations should not be computationally heavy, otherwise, it would cancel the benefit of the reduced response time achieved through the reduction in data size.

*Example.* The following two implementations found in the reviewed papers demonstrate the Feature Selection and Extraction tactic.

- Cloud-based Threat Detector [S10] collects and analyze various types of data (e.g., system logs, firewall logs, and router logs) to detect cyber attacks such as DDoS attack and port scanning attack. This system leverages Naïve Bayes algorithm to select five features out of 50 features in CAIDA dataset. These features include source IP address, destination IP address, port, packet length, and protocol. After selecting the features, parallel processing framework (i.e., MapReduce) is used to extract the features from the records. The incorporation of Feature Selection and Extraction tactic reduces the size of dataset from 200 GB to 50 GB. It was found that incorporation of this tactic does not have any negative impact on the detection accuracy of a system.
- Quasi Real-Time IDS [S12] leverages Information Gain Ranking Algorithm to select eight features from network traffic records in CAIDA dataset. After selecting the features, parallel processing framework (i.e., MapReduce) is used to extract the features from the records. A distinguishing characteristic of this system is that it also allows the user to select features manually at runtime. This flexibility is helpful in situations where the dynamics of network traffic changes frequently. This is enabled by Apache Hive[5] and Tshark[51]. The Feature Extraction Perl Script enables a user to select the features from a record using Tshark and then creates a table in Hive accordingly.

*Dependencies.* The Feature Selection and Extraction tactic requires Parallel Processing tactic (Section 5.1.4) to speed up the process of feature selection and extraction. It is worth noting that Unnecessary Data Removal tactic (Section 5.1.2) and Feature Selection and Extraction tactic are not interchangeable. Both of these tactics play separate roles in improving the performance of the security analytic system.

### 5.1.4. Parallel processing

*Introduction.* The Parallel Processing tactic can be found in all (i.e., 74) of the reviewed studies. This tactic distributes the processing of a large amount of security event data among different nodes of a computing cluster. The nodes process the data in parallel fashion, which significantly improves the response time of a system.

*Motivation.* There are a number of sources within an enterprise that generate security event data. These sources include but not limited to network devices (e.g., switches and routers), database activities, application data, and user activities. Security event data is generated at a very high speed. For example, an enterprise as large as HP used to generate around one trillion security events per day in 2013, which was expected to grow further in the coming years [4]. A standalone computer that processes such a large size of security event data in a sequential manner will take a lot of time to detect an attack, which is not tolerable in such security-critical situations.

*Description.* Fig. 10 shows the main components of the Parallel Processing tactic. The numbers in the figure show the sequence of operations. The *data collector* component collects security event data from different sources depending on the type of security analytics and security requirements of an enterprise. The *data collector* forwards the collected data to *data storage* component, which stores the data. Data can be stored in several ways such as Hadoop Distributed File System (HDFS), HBase, and Relational Database Management System (RDBMS). In order to enable parallel processing, the stored data needs to be partitioned into fixed-size blocks (e.g., 64MB or 128MB). After partitioning, data is processed in the *data analysis* component through several nodes working in parallel according to the guidelines of a distributed framework such as Hadoop or Spark. The result of analysis is shared with the user through the *visualization* component.

*Constraints.* The Parallel Processing tactic assumes that a security analytic system incorporating this tactic is already integrated with a cluster of nodes capable of processing data in a parallel fashion. Another important factor that needs to be taken care of is the breaking of a logical record across two blocks during the partitioning of data into blocks. In such a situation, it is important to keep enough information about the file data type so that record can be reconstructed.

---

[5] https://hive.apache.org/

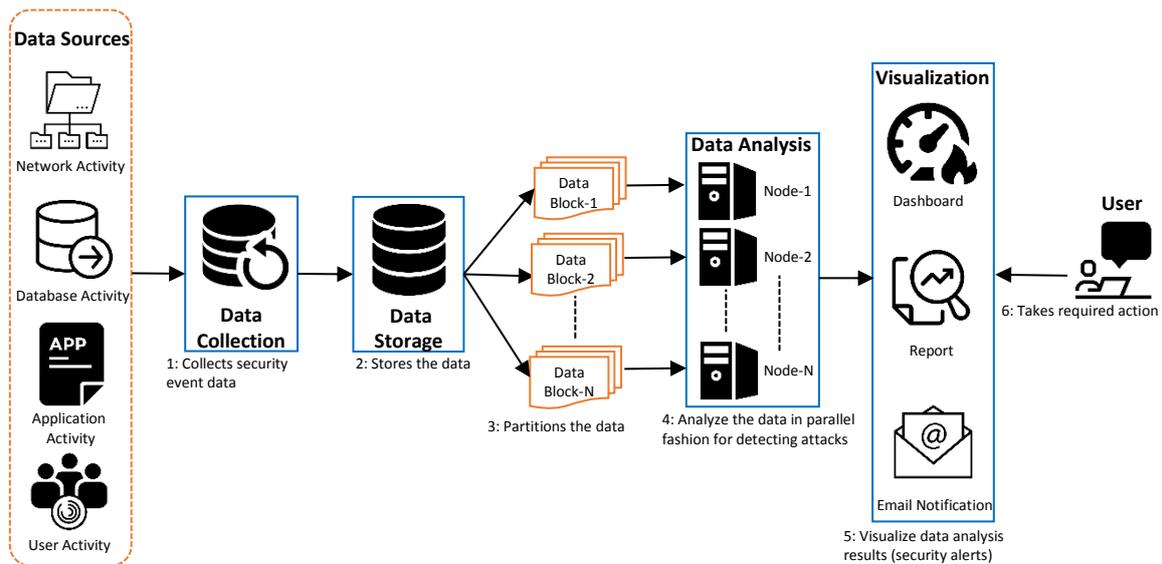

Fig. 10. Parallel processing tactic

*Example.* Honeypot-based Phishing Detection [S17] demonstrates the applicability of this tactic for improving the response time of a phishing attack detection system. The authors compared the response time of a sequentially implemented system with a parallel-implemented system, each processing 268 GB of security event data. It was found that sequentially implemented system took 180 minutes to process all data. On the other hand, parallel implementation of a system with Hadoop and Spark frameworks took 21 minutes and 14 minutes respectively with a cluster of 9 nodes. The authors also demonstrated that the more the number of nodes in a parallel processing scenario, the faster will be the response time of the system. For example, the response time of Hadoop was recorded as 57, 36, and 21 minutes with 3, 5, and 9 nodes respectively.

*Dependencies.* The Parallel Processing tactic depends upon the Dynamic Load Balancing tactic (Section 5.3.1) and the Data Ingestion Monitoring tactic (Section 5.4.1) for balancing the load among the nodes and controlling the flow of the data into the nodes respectively.

**5.1.5. Result polling and optimized notification**

*Introduction.* Result Polling and Optimized Notification tactic is found in Count Me In [S56]. This tactic helps optimize the delay caused due to a predefined time interval for feeding the results from the mapper nodes into the reducer nodes inside a parallel processing security analytic system. This tactic ensures that as soon as values inside the mapper nodes change to a sufficient degree (set by the admin), the mapper node notifies the reducer node and accordingly forward the updated results to the reducer node.

*Motivation.* MapReduce is a parallel processing framework that is widely adopted in a distributed setup [52]. This framework consists of two phases – Map and Reduce. In the Map phase, features derived from the network traffic maps to a key and a value through multiple mapper nodes. For example, a key-value pair (s, d) may represent failed connection attempts from a source address '*s*' to a destination address '*d*'. In the Reduce phase, the key-value pairs generated by mapper nodes are fed into multiple reducer nodes for generating results. The generated results are evaluated against a predicate threshold through a trigger and if the results exceed a specific limit, then an alert is generated that signal towards a possible cyber attack.

The key-value pairs from mapper nodes are fed into reducer nodes after a predefined time interval (e.g., 5-minute aggregation or 1-hour aggregation). Once reducer generates the results, the trigger executes the predicate threshold. This predefined time interval introduces a delay. For example, an attack may be launched at t=5 sec of the predefined time interval and the trigger will be executed at t=5 min. In this 5 min, it is quite possible that a significant damage might already have been caused.

*Description.* The incorporation of Result Polling and Optimized Notification tactic in a security analytic system is demonstrated in Fig. 11. *Data collection* component collects security event data from multiple sources. The collected data is stored by the *data storage* module. Next, *parallel data processing* component reads the stored data. The *parallel data processing* component consists of mappers and reducers. The number of mappers and reducers depends upon the number of nodes in the underlying cluster. The mapper sub-components (e.g., Mapper Node-1 and Mapper Node-2) read the data in parallel and produce intermediate key-value pairs shown as the result in the Fig. 11. When all the map jobs are completed, the key-value pairs (shown as values in the figure) are passed on to the reducers. The reducers merge the values to produce the final results. Normally, the reducers have to wait until all map jobs are completed, which causes a significant delay in responding to cyber attacks. To address this issue, the mechanism of the poll and notify is introduced. According to this mechanism, the mappers monitor the changes in incoming security event data. As soon as the change in the security event data crosses a predefined threshold, the mappers notify the reducers to get ready for taking the required data without waiting for completion of the predefined time interval. In addition to the optimized notification from mappers, the reducers can poll the mappers for results depending upon the processing capacity of the reducer(s). After receiving the intermediate key-value pairs from mappers, the reducers process the intermediate results to produce the final results. Immediately after the reducers produce the final results, the trigger is executed to check for possible cyber attacks.

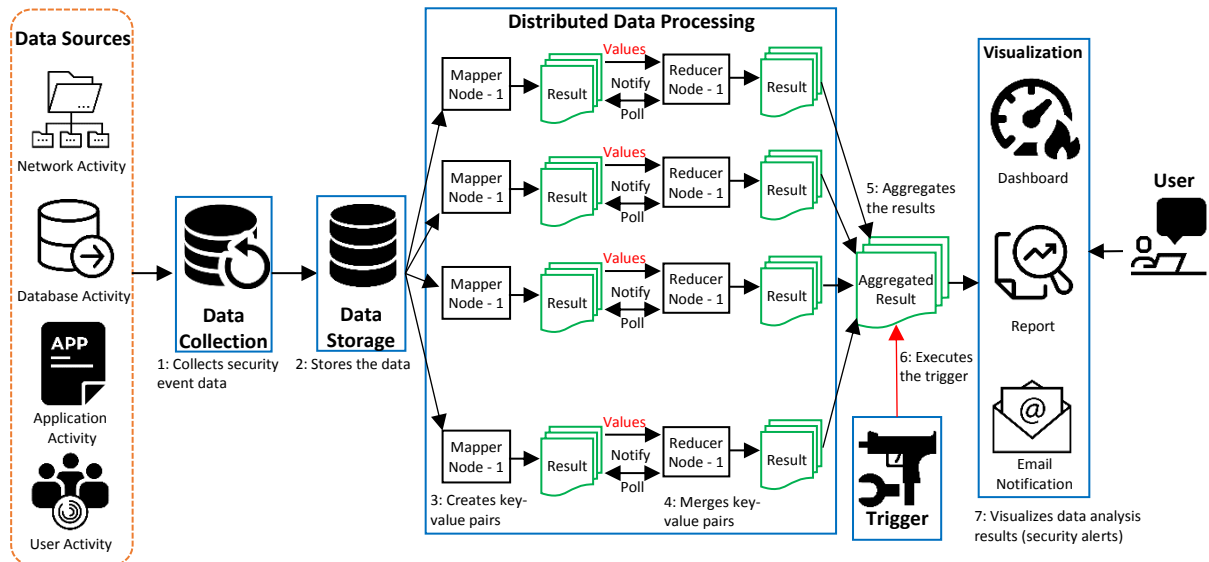

Fig. 11. Result polling tactic and optimized notification tactic

*Constraints.* In a distributed setup, it is important that the threshold set for change in intermediate values at mapper nodes is applied to the average of all mapper nodes. For example, if a change in values at a single mapper node crosses the defined threshold, but the rest of the mapper nodes are not experiencing sufficient change in values, intermediate results should not be passed to the reducer nodes. This is due to the reason that by-passing the defined time interval and forwarding unchanged intermediate results from mappers to reducers will put extra processing burden on reducers without any significant gain.

*Example.* Count Me In [S56] experimentally examines the communication overhead caused by the incorporation of this tactic. The system monitors an uplink traffic averaging 1 Gb/sec during day-time hours. It is observed that out of the total messages communicated among the nodes of a system, only 0.40% relates to the Result Polling and Optimized Notification tactic. In total out of 69, 810 notifications from mapper node to reducer nodes, the reducer nodes ignore 27, 704 notifications to limit simultaneous outstanding key updates.

*Dependencies.* The Result Polling and Optimized Notification tactic requires Parallel Processing tactic (Section 5.1.4) to enable the reducer nodes to share the extra load generated due to result polling among the nodes. This tactic also depends upon the MapReduce tactic (Section 5.3.2) for providing the programming framework to design mappers and reducers.

*Variations: User-guided poll and notify.* The Result Polling and Optimized Notification tactic as described requires either the mappers to notify the reducers about updates or requires reducers to poll the mappers for updates. However, in some systems, the task of poll and notify can be relegated to a user. A user can initiate the process of sending intermediate key-value pairs from mappers to reducers at any point of time irrespective of the predefined time interval. For example, SEAS-MR [S38] has both the options – periodic aggregation and user-guided aggregation of intermediate results produced by mappers. The user-guided poll and notify can also be used for long-term analysis where a user can specify the start time and the end time for an interval. In this system, periodic aggregation is implemented with MapReduce as it is better for performance while user-guide aggregation is implemented with Pig[6] script as it facilitates user interactivity.

#### 5.1.6. Data Cutoff

*Introduction.* The Data Cutoff tactic has been identified from Forensic Analyzer [S30] and VALKYRIE [S45]. This tactic applies a customizable cutoff limit on each network connection or process to select and store data pertaining to a specific portion of the connection or process. For example, selecting and storing only first 15 KB of network traffic data for a connection or data pertaining to first 100 sec of the execution time of a process. Such a cutoff reduces size of the dataset for security analysis, which helps improve the overall performance of the system.

*Motivation.* Due to the ever-increasing volume of security relevant data (e.g., network traffic, system logs, and application activity), it is infeasible to collect, store, and analyze the data in its entirety. For example, Lawrence Berkeley National Laboratory (LBNL), a security research lab containing around 10,000 hosts, experiences around 1.5 TB of network traffic per day. In majority of the cases, only a small subset of the security event data turns out to be relevant for security analysis [24]. In case of network connections, more connections are short with few large connections accounting for bulk of total volume [25, 26]. Thus, by selecting and storing the first N bytes (cutoff) for each large connection, the connection can be stored in its entirety. This is because the beginning of such connections is the portion of interest that contains information such as protocol handshakes, authentication logs, and data item names.

*Description.* Fig. 12 shows the main elements of Data Cutoff tactic with the numbers to indicate the sequence of the operations. The *data collection* module collects security event data from one or several available sources and passes the collected data to *data cutOff* module. Examples of security events include network security event (e.g., [source IP, destination IP, port, protocol]) and process event (e.g., [file name information, privilege level, parent process ID, timestamp]). The *data cutOff* component enforces the cutoff by discarding security events that appear after a network connection or process has reached its predefined limit. Any security event that appears after the predefined limit does not contribute significantly to the attack detection process, therefore, analyzing such security events put an extra burden on data processing resources without any significant gain. The security event data left after cutoff is stored by the *data storage* component. The stored data is read by *data analysis* module to analyze it for detecting cyber attacks. Finally, the result of analysis is displayed to a user through *visualization* component. A user takes the required action upon arrival of any outstanding alerts.

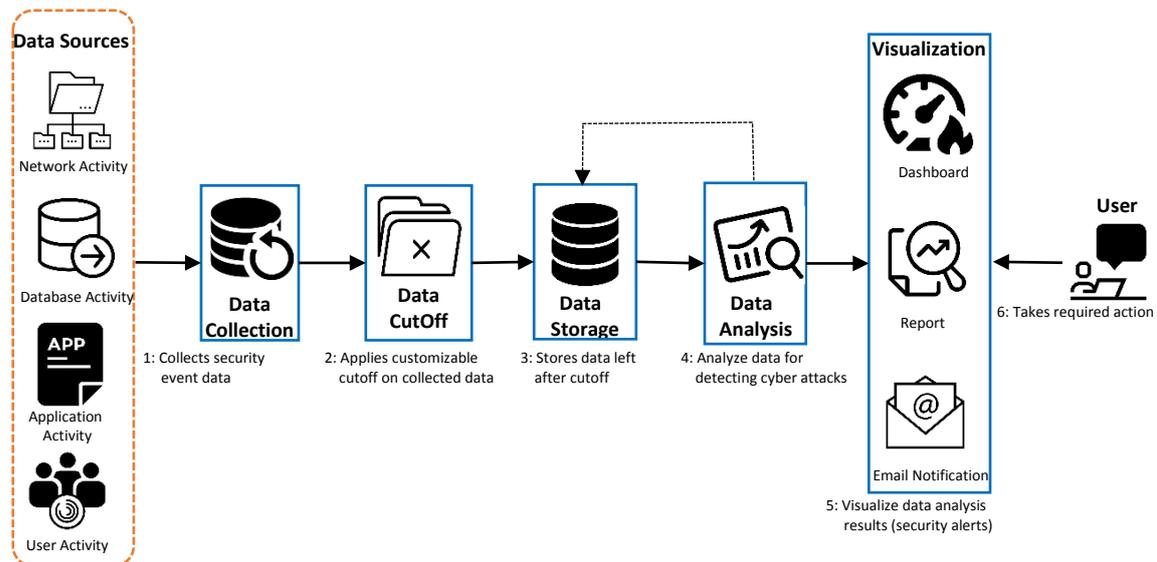

Fig. 12. Data cutoff tactic

*Constraints.* The Data Cutoff tactic poses a risk of evading an attack from detection. If an attacker is smart enough to initiate an attack after the cutoff limit, a security analytic system will not record the malicious activity, and so the attack will go undetected. There are several delaying tactics usually employed by attackers with the intent to evade such tactics that record only initial portion of an interaction. Several compensation techniques can be employed to reduce the risk of such an evasion. These techniques include: (1) selecting cutoff limit according to the type of data sources as delaying an attack to later stages is harder for some services as compared to others (2) the cutoff limit can be increased for data sources where there is a higher risk of an attack (3) the application of cutoff can be randomized instead of always being applied at the start of an interaction (e.g., network connection or a process) so that an attacker cannot predict at which point the cutoff will come into play. In addition, the functionality required for implementing Data CutOff tactic should not be computationally expensive, otherwise, it will cancel the benefit that is expected through the incorporation of this tactic.

---

[6] https://pig.apache.org/

*Example.* The Data CutOff tactic has been implemented in Forensic Analyzer [S30] and VALKYRIE [S45].

- Forensic Analyzer [S30] is a cloud-based network security forensic analysis system that is evaluated by detecting phishing attacks based on analysis of captured network traffic. The data collected in a span of six months is around 20 TB in size. The Data CutOff tactic was applied to the collected data with a cutoff limit of 15 KB. This means to only select the first 15 KB of each network connection. The incorporation of this tactic reduces the size of data from 20 TB to 1 TB. The effect of data cutoff on detection accuracy of a system has not been reported.
- VALKYRIE [S45] is a security analytic system that detects malware attacks based on an analysis of kernel-level telemetry data. The system implements Data CutOff tactic to reduce the size of data by selecting the data pertaining to only first 100 seconds of the execution time of a process. Despite the incorporation of Data CutOff tactic, the system achieves detection accuracy of around 97-99%.

*Dependencies.* The Data CutOff tactic requires Parallel Processing tactic (Section 5.1.4) to speed up the process of discarding security event data that appears after the predefined cutoff limit.

5.2. Accuracy

This section reports the architectural tactics that are related to Accuracy quality attribute.

**5.2.1. Alert Correlation**

*Introduction.* Alert Correlation tactic have been found in [S6], [S7], [S11], [S16], [S19], [S38], [S39], [S41] and [S65]. This tactic analyzes the individual alerts produced by security system(s), discards irrelevant alerts, and groups together the relevant alerts based on a logical relationship between them to provide a global and condensed view of the security situation of an organization's infrastructure (i.e., network and hosts). The incorporation of this tactic improves the accuracy of a security analytic system by reducing the number of false positives and detecting highly sophisticated and complex attacks.

*Motivation.* Organizations employ different security tools and technologies for better detection coverage of their networks and hosts. For example, a typical organization may deploy firewall, anti-virus, and IDS for accepting/dropping network traffic, scanning malware based on predefined signature, and detecting known attack patterns or abnormal behaviours respectively. Unfortunately, these security tools generate a large number of alerts. For instance, an IDS deployed in a real-world network generates around 9 million alerts per day [56]. Investigating and responding to these many alerts is quite challenging especially when 99% of them are false positives [57]. Furthermore, these security tools monitor the security in isolation without considering the context and logical relationship between alerts generated by other security tools. This isolated approach is not able to detect attacks that operate in slow mode over a period of time where some alerts are precursors to more complex and dangerous attacks. To address these issue, a high-level management is required that correlate the alerts by taking context and their logical relationship into consideration before reporting them to users.

*Description.* Fig. 13 shows the main components involved in the Alert Correlation tactic. The *data collection* module collects security event data from different sources. The collected data is stored in the *data storage* and copied to the *data pre-processor* component for pre-processing the raw data. The pre-processed data is ingested into the *alert analysis* component, which analyzes the data for detecting attacks. It is worth mentioning that Alert analysis component analyzes the data in an isolated fashion (without considering any contextual information) either using misuse-based analysis or anomaly-based or both. The generated alerts are transferred to the *alert verification* component, which uses different techniques [58] to determine whether an alert is a false positive. The alerts identified as false positives are discarded at this stage. The clean and synthesized alerts are forwarded to the *alert correlation* component for further analysis. The alerts are correlated (i.e., logically linked) using different techniques [58] such as scenario-based correlation, rule-based correlation, statistical correlation, and temporal correlation. The Alert correlation component coordinates with *data storage* for taking the required contextual information about alerts. The results of correlation are released through the *visualization* component. Finally, either an automated response is generated, or a security administrator analyzes the threat and responds accordingly.

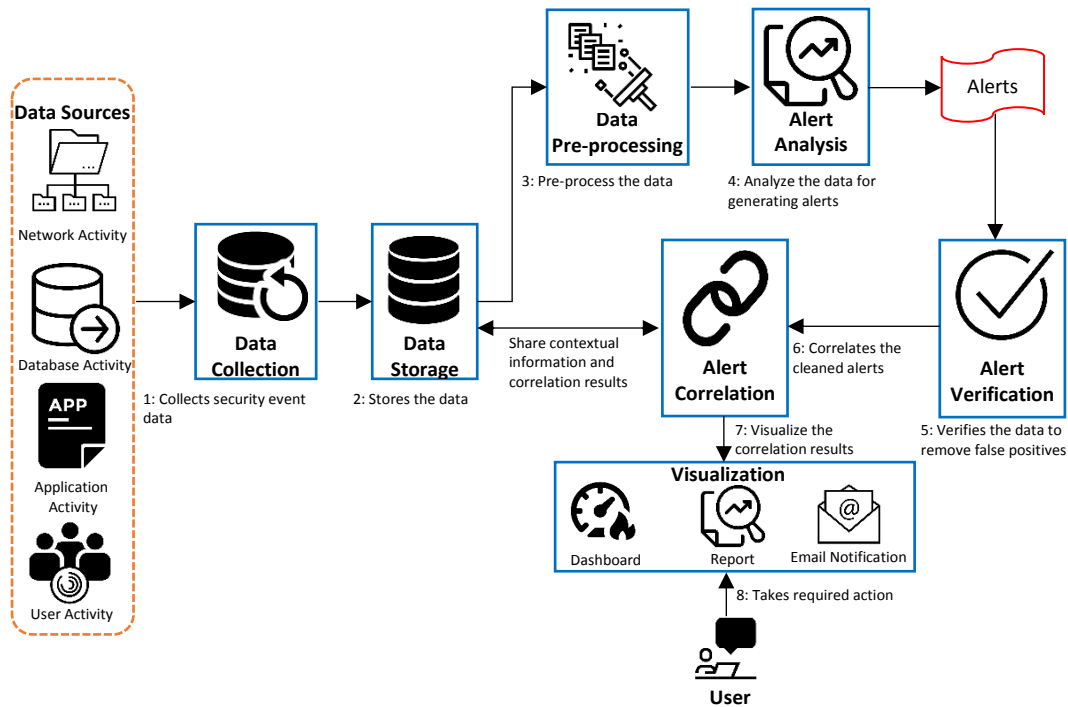

Fig. 13. Alert correlation tactic.

*Constraints.* The in-depth analysis employed for alert correlation improves the accuracy but increases the response time due to the inclusion of an extra complex computation stage to a system. The tactic also requires mechanisms for acquisition of domain knowledge and adaptation to changes in the networks and host infrastructure [58].

*Examples.* The nine systems that implement the Alert Correlation tactic collect individual alerts from different security systems (e.g., IDS or Firewall) and correlate them for detecting attacks. What varies between these nine systems is the correlation method employed. We outline the underlying correlation techniques for a few of them.

- GSLAC [S7]: The system employs causal-based technique for correlating the alerts. Each alert is treated as a vector with multiple attributes (i.e., destination IP, source IP, timestamp and so on). The alerts are presented in the form of a graph where each alert has a prerequisite alert and a consequence. A security analyst analyzes the graph for identifying complex attack scenarios.
- Hunting attacks in the dark [S41]: This system correlates the alerts based on their source and destination IP addresses. The similarity between IP addresses for different alerts is measured using intersection cardinalities and if the similarity score is above a predefined threshold, it means alerts belong to same IPs that signals towards a potential attack.
- Multistage alert correlator [S6]: This system employs a model of prerequisites and consequences proposed in [59] for determining the relationship among individual alerts. The alerts are correlated based on the similarities among their source IP, destination IP, start time, and end time. A graph is generated that shows each alert with its prerequisite and consequence. If a prerequisite of an alert is present in the graph as a consequence of a previous alert, the two alerts are closely related and analyzed for picturizing the complex attack scenario.

*Dependencies.* The Alert Correlation tactic will correlate alerts of any quality; however, effective correlation requires the alerts to be of a good quality that is dependent upon the tactics employed in the data analysis module such as Attack Algorithm Selection tactic (Section 5.2.3) and Combining Signature-based and Anomaly-based Detection (Section 5.2.2).

### 5.2.2. Combining signature-based and anomaly-based detection

*Introduction.* Genetic Algo-based Distributed Denial of Service (DDoD) Detection [S61], Hybrid Intrusion Detection [S72], and Snort + PHAD + NETAD [60] employ this tactic. The Combining Signature-based and Anomaly-based Detection tactic enables a security analytic system to analyze the collected security event data for two objectives: (1) find a match with the already available attack patterns or signatures and (2) find a deviation from the learned normal behaviour i.e., an anomaly. In either case, finding a match or a deviation, a system generates an alert signalling towards a possible cyber attack. The combination of misuse and anomaly significantly improves the detection accuracy and reduces the false positive rate.

*Motivation.* Based on detection principle, security analytic systems are of two types – Signature-based (often called misuse-based) and Anomaly-based. Signature-based systems detect attacks based on predefined attack patterns. These patterns are designed based on already reported attacks. If an ongoing activity matches with the attack pattern, the activity is termed as malicious. These types of systems are very effective in detecting known attacks but are unable to detect unknown attacks [61]. Anomaly-based systems learn the normal behaviour of an organization's infrastructure and any activity that deviates from this behaviour is termed an attack. This class of systems can detect unknown attacks, however, it generates a large number of false positive alarms [62]. Considering the limitations of both type of systems, it is important to come up with a solution that can minimize these limitations.

*Description.* The main components of the Combining Signature-based and Anomaly-based Detection tactic are shown in Fig. 14. The *data collection* component collects security relevant data from different sources. The collected data is stored by the *data storage* component. Next, data is fed into the *signature-based detection* module that analyzes the data to identify attack patterns. For such analysis, this component leverages the pre-designed rules from the *rules database* that define attack patterns. If a match is identified, an alert is directly generated through a *visualization* component. If *signature-based detection* module does not detect any attack pattern in the data, the data is forwarded to the *anomaly-based detection* module for detecting unknown attacks that cannot be detected by the *signature-based detection* module. The *anomaly-based detection* component analyzes the data using machine learning algorithms to detect deviations from the normal behaviour. When an anomaly (deviation) is detected, an alert is generated through the *visualization* component. At the same time, the *anomaly* is defined in the form of a *rule* or attack pattern and added to the *rules database*. This way the *rules database* is constantly updated to enable the *signature-based detection* module to detect a variety of attacks.

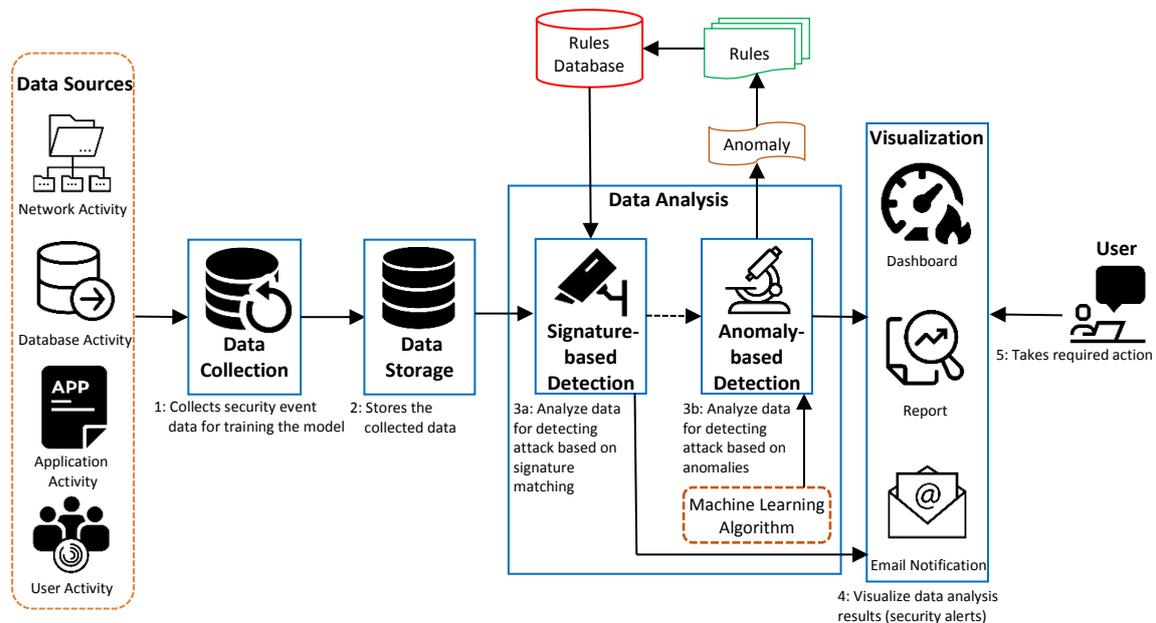

Fig. 14. Combining signature-based and anomaly-based detection tactic

*Constraints.* Combining Signature-based and Anomaly-based Detection tactic may impact the overall performance of a system by introducing additional data analytic requirements. Furthermore, combining signature-based and anomaly-based detections into a single security analytic system and getting them interoperate in an efficient and successful way can be more complex and challenging.

*Example.* The following two systems demonstrates the implementation of the Combining Signature-based and Anomaly-based Detection tactic.

- Hybrid Intrusion Detection [S72]: This system combines signature-based IDS (Snort) with an anomaly-based detector implemented using Hadoop and Hive. The hybrid system is compared with the stand-alone signature-based system in terms of accuracy. Both the systems have been evaluated with several attacks such as ICMP, Smurf, SYN flood, UPD, and Port scanning attack. It is found that a hybrid system achieves higher detection rate for all attacks as compared to a stand-alone system. For example, for port scanning attack, a stand-alone system achieves a detection rate of around 40% while a hybrid system shows a detection rate of around 50%.
- Snort + PHAD + NETAD [60]: This system combines signature-based IDS (Snort) with two anomaly detectors i.e., Packet Header Anomaly Detector (PHAD) and Network Traffic Anomaly Detector (NETAD). The hybrid system is evaluated using IDEVAL dataset [63] to explore its detection rate in comparison to the stand-alone signature-based system. It is observed that the stand-alone system is able to detect only 27 attacks while the hybrid system (i.e., Snort+PHAD+NETAD) detects 146 attacks from the dataset.

*Dependencies.* The Combining Signature-based and Anomaly-based Detection tactic requires Parallel Processing tactic (Section 5.1.4) to reduce the additional analysis time required due to two-phase analysis (i.e., signature-based and anomaly-based). Additionally, this tactic also depends upon Attack Detection Algorithm Selection tactic (Section 5.2.3) to efficiently select an effective algorithm for detecting anomalies in the security event data.

**5.2.3. Attack detection algorithm selection tactic**

*Introduction.* We have already stated that all of the reviewed systems leverage some type of machine learning algorithm to analyze the security event data for detecting cyber attacks, therefore, this tactic can be found in almost all of the reviewed systems. Unlike ML Algorithm Optimization tactic (Section 5.1.1) that emphasizes the role of ML algorithm on performance, the objective of Attack Detection Algorithm Selection tactic is to highlight the role of the algorithm in improving the accuracy of a security analytic system and provide some guidelines for selecting ML algorithm that is most appropriate for accurate detection of attacks.

*Motivation.* Cybersecurity analytics systems are expected to accurately detect cyber attacks without generating security alerts for legitimate activities (i.e., false positives). In addition to several other factors such as data source and data quality, the employed machine learning algorithm plays a significant role in accurately detecting cyber attacks. The security analytics systems use machine learning algorithms to classify the security event data either as legitimate (corresponding to normal activity) or malicious (corresponding to attacks). There exists a variety of machine learning algorithms that can be leveraged to detect attacks. These algorithms include but not limited to Logistic Regression, Support Vector Machine, Naïve Bayes, Random Forest, and Gradient Boosted Decision Tree, K-means, and Artificial Neural Networks. For details on machine learning algorithms used in cybersecurity analytics, readers should refer to [53]. The big challenge here is to determine which of the many available algorithms will yield the most accurate results.

*Description.* The major components of Attack Detection Algorithm Selection tactic are shown in Fig. 15. The *data collection* component collects security event data for training the security analytic system for detecting cyber attacks. The training data can be collected from sources within an enterprise where a system is supposed to be deployed as depicted in Fig. 15 or an already available popular dataset such as KDDCup99 can be used as training dataset. How much data should be sufficient to train the model varies from one case to another. For example, the famous DARPA 1998 dataset contains data collected from network and operating system in a span of nine weeks. The first seven-weeks data were assigned as training data and the last two weeks as testing data. After collecting the training data, the *data preparation* component prepares the data for training the model by applying filters and feature extraction techniques. Next, the prepared training data start training the attack detection model. Once the model is trained, it is tested to investigate whether the model can detect cyber attacks. For testing the model, the data is collected from an enterprise as shown in step 4. The test data is prepared for feeding into the attack detection model. The prepared test data is forwarded to the *attack detection model*, which analyzes the data based on the rules learned during the training phase. Here, the incoming test data instances are classified as either legitimate or malicious. The analysis results are displayed to a user through the *visualization* component. In case of malicious or attack situation, a user can take immediate actions that may include blocking certain ports or cutting off the affected components from the network to stop further damage.

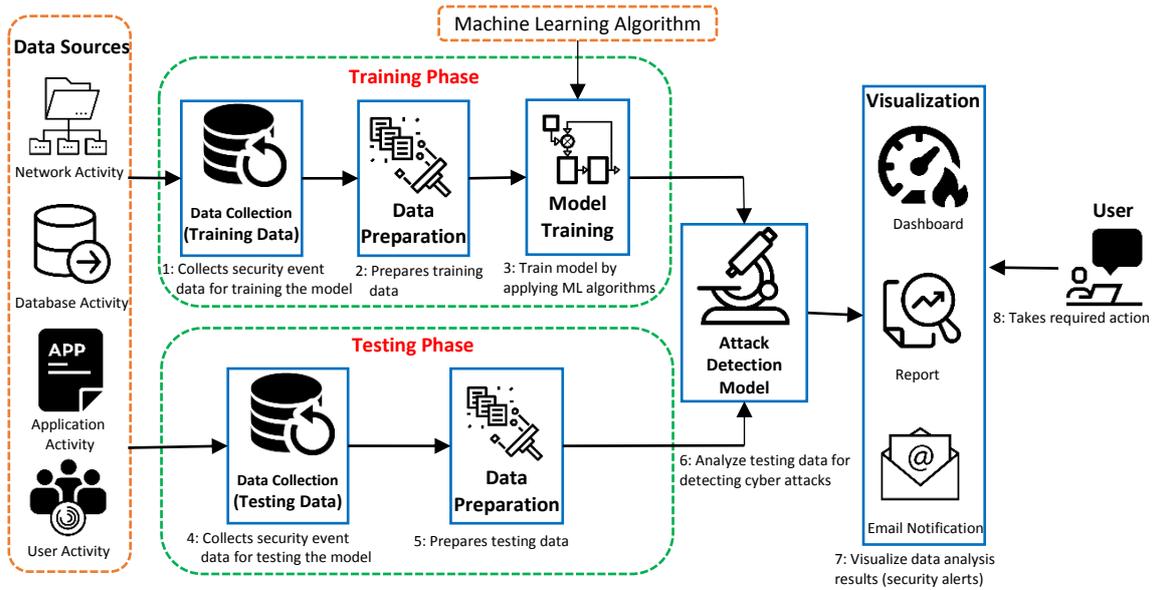
Fig. 15. Attack detection algorithm selection tactic

*Constraints.* While selecting an attack detection algorithm, software architect should keep an eye on several things due to the following reasons

- The selection of ML algorithm is tricky because an algorithm may work well for detecting one type of attack but may not work well for detecting other types of attacks
- Similarly, algorithm selection is challenging in the sense that in addition to accuracy, it affects other system's qualities such as response time, complexity, and understandability of the final result. For example, Cheng et al. [55] compare SVM with Extreme Learning Machine (ELM) in terms of accuracy and performance. It has been revealed that SVM generates more accurate results but is computationally expensive. On the other hand, ELM generates less accurate results, but is more light-weight. A reasonable trade-off should be established among various system's qualities while selecting an algorithm.
- There is no such standard available that compares the accuracy of all ML algorithms on the same dataset. Different research explorations use different datasets (e.g., KDD99 and DARPA 1999) or different subsets from the same dataset, which makes it harder to fully rely on these findings for selecting an algorithm for a specific security analytic system.
- The applicability of the ML algorithm closely relates to the quality of the data. Therefore, software architect needs to carefully asses the quality of the available data and then accordingly select the attack detection algorithm.
- The selection of algorithm also depends upon the working mode of a security analytic system [53]. An algorithm that best fits for offline analysis may not be suitable for online analysis.

*Example.* Whist all of the reviewed systems leverage ML algorithms, we report only those systems that explore and compare the effects of multiple algorithms on the attack detection rate of a system. The objective is to exhibit the importance of algorithm selection for the detection rate of a security analytic system.

- Streaming-based Threat Detection [S1]: This system explores the applicability of K-means clustering algorithm [64] and Fuzzy C-means clustering algorithm [65] for accurately detecting cyber attacks. The system is evaluated to detect DDoS and flooding attacks based on analysing 260 GB of network traffic data gathered at Chicago Equinix data centre [50]. Experimental results show that detection rate for K-means is 91.8% with 1.8% false positive while for Fuzzy C-means the detection rate is 86.5% with 2.7% false positive.
- Improved K-means IDS [S27]: K-means algorithm is improved by specifying criteria for selecting initial centre of cluster which is random in traditional K-means algorithm. Both K-means and improved K-means are implemented in IDS for comparing detection accuracy. It has been found that improved K-means accurately detects 89% of the attacks with 2.4% false positive in KDDCup99 dataset. On the other hand, K-means shows an accuracy of 86% with 1.0% false positive rate.

- Spark-based IDS Framework [S9]: This framework compares five machine learning algorithms: Logistic regression, Support vector machine, Random forest, Gradient boosted decision trees, and Naïve Bayes. Two datasets, KDDCup99 and NSL-KDD, are used to evaluate the accuracy of the system with each of the algorithms. With KDDCup99, Logistic regression shows the best accuracy (i.e., 91%) and SVM shows the worst (i.e., 78%). With NSL-KDD, the best accuracy is achieved with Random Forest (i.e., 82.3%) while SVM delivered the worst accuracy rate of 37.8%.

*Dependencies.* The Attack Detection Algorithm Selection tactic depends upon Unnecessary Data Removal tactic (Section 5.1.2) and Feature Selection and Extraction tactic (Section 5.1.3) to help bring the collected data into a refined form. After the application of Unnecessary Data Removal and Feature Selection and Extraction tactic, attack detection algorithm can be efficiently applied to the refined data to accurately detect cyber attacks. The Attack Detection Algorithm Selection tactic needs to be incorporated alongside ML Algorithm Optimization tactic (Section 5.1.1) as these two tactics establish the trade-off between the effects of the ML algorithm on accuracy and performance.

**5.2.4. Combining multiple detection methods**

*Introduction.* The Combining Multiple Detection Methods tactic has been used in Multiple Detection Methods [S46]. This tactic applies and integrates the results of multiple security analytic methods on security event data. The incorporation of multiple security analytic methods in a security analytic system reduces false positive alarm rate and improves the attack detection accuracy.

*Motivation.* Most of the security analytic systems employ a single detection principle for detecting attacks. Examples of detection principles include traffic volume to a server crosses a predefined threshold, a large number of port access in a certain period of time, and IP address that generate traffic above a certain threshold. The disadvantage of employing a single detection principle is that the system generates a large number of false positives [S46]. Analysing the security-relevant data through multiple detection principles reduces the number of false positives by integrating the results of multiple detectors. Each detector accommodates a different detection principle. For example, SYN flood attack can be detected by monitoring or analysing the number of network flows in network traffic data while a DoS attack can be detected by monitoring the volume of network traffic between source and destination IPs.

*Description.* The main components of the Combining Multiple Detection Methods tactic are shown in Fig. 16. Security event data is collected from different sources. It is worth noting that the sources from where security event data can be collected is not limited to what is shown in the figure. The choice of data sources varies from organization to organization depending upon their specific security requirements. After collection, data is stored in a *data storage*. Then the data is forwarded to the *data analysis* module where various attack detection methods are applied to analyze the data. The number and choices of the attack detection methods depend upon several factors. These factors include the processing capability of an organization, the data sources, security requirements, and finally the security expertise of an organization. For example, a highly security-sensitive organization (e.g., National Security Agency) with high budget and computational power may incorporate more number of attack detection methods to secure their data and infrastructure from cyber attacks. The attack detection methods are applied to the whole dataset in a parallel fashion. The *visualization* module immediately reports any outstanding anomalies to users or administrators, who are expected to respond to security alerts.

*Constraints.* Combining Multiple Detection Methods tactic requires more computational power as compared to a security analytic system that incorporates a single detection method. The more the number of detection methods employed by a system, the more computation power will be required. In other words, this tactic will affect the overall performance of the system.

*Example.* The Combining Multiple Detection Methods tactic has been implemented in Multiple Detection Methods [S46]. This system adopts three detection methods for analyzing network traffic data. These methods include (1) Intuitive methods based on network flows, (2) Intuitive methods based on traffic volume, and (3) Least-Mean-Square(LMS)-based detection methods. Intuitive methods based on network flows detect attack based on the number of network flows, intuitive methods based on traffic volume detect attacks based on the number of network packets, and LMS-based detection methods detect attack based on the correlation between the number of IP flows and traffic volume. The system is tested with two types of attacks – SYN flooding attack (a large number of small network flows) and DoS attack (a small number of large network flows). The security analytic system detects SYN flooding attack through an intuitive method based on network flows as the number of network flows during the attack crosses the threshold. On the other hand, DoS attack is detected through an intuitive method based on traffic volume as the number of packets in a particular network flow when an attack exceeds the predefined threshold.

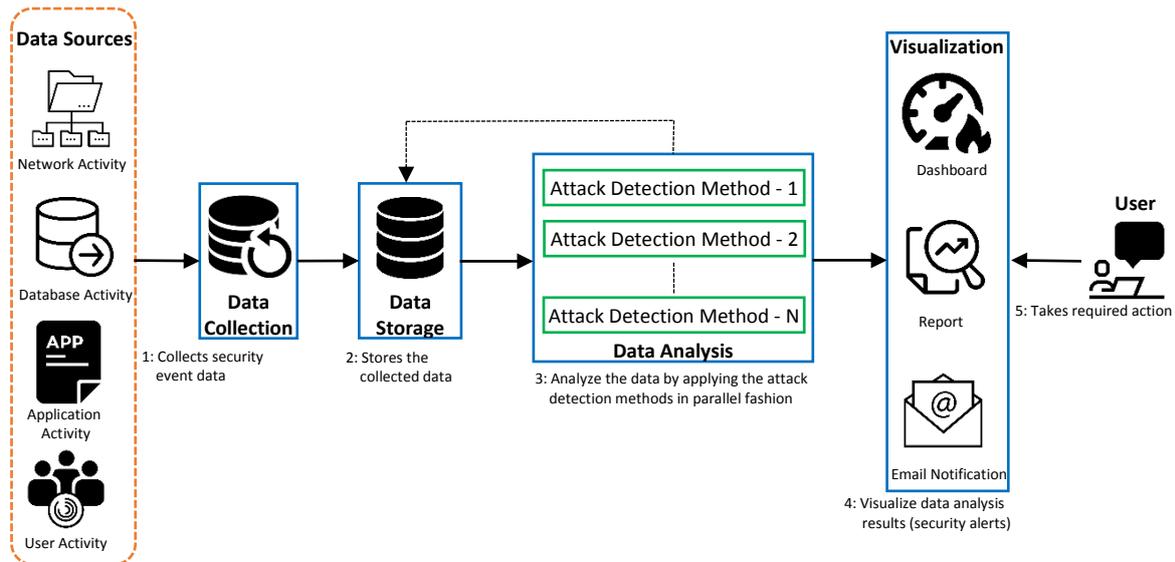

Fig. 16. Combining multiple detection methods tactic

*Dependencies.* The Combining Multiple Detection Methods tactic depends upon Parallel Processing tactic (Section 5.1.4) that can provide the processing speed required for applying multiple attack detection methods on security event data. Without parallel processing, the system's response would be too slow.

5.3. Scalability

This section reports the architectural tactics that are related to Scalability quality attribute.

**5.3.1. Dynamic load balancing**

*Introduction.* Dynamic Load Balancing tactic can be found in GSLAC [S7] and Cloud Bursting [S34]. This tactic is used to balance the processing load among analysis nodes by dividing the security event data among the nodes. Having processing capacity available in a cluster, the Dynamic Load Balancing tactic makes a system scale well without adding any further hardware resources.

*Motivation.* A security analytic system leverages a cluster of computing nodes for storage and analysis of security event data. The size of the cluster is different in different systems (e.g., 10 nodes in [S7] and 5 in [S34]). The system distributes the security event data among the nodes to speed up the process. When the speed of data input increases (for instance from 100 MB/sec in weekdays to 150 MB/sec over weekends), it is important that the system distributes the increased load in a balanced way to avoid a situation where one node is under extreme load (i.e., 100% CPU utilization) and another node is under-loaded (i.e., 30% CPU utilization).

*Description.* Fig. 17 shows the main components of the Dynamic Load Balancing tactic. The *data collection* component collects data from different sources. The captured data is sent to the *data filtration* component, which removes the data that does not contribute to the attack detection process. The filtered data is forwarded to a *load balancer*, which distributes the data among different nodes to balance the workload among the nodes in the *data analysis* module. Data can be distributed based on different criteria. For example, network traffic can be distributed based on header information (e.g., IP address or TCP ports) or payload information (e.g., signatures). The balanced distribution of data is not fully reliable due to the heterogeneous nature of the security event data [66, 67]. For example, if network traffic is distributed based on IP range, one range may contain more number of packets with large size, which may exhaust one node. On the other hand, a node handling another IP range with few packets of small size might be sitting idle. Therefore, the *resource monitor* component is introduced that constantly monitors the CPU utilization of nodes and reports to the *load balancer*. When the difference between the CPU utilization of the nodes crosses a predefined threshold, the *load balancer* rebalances the load among the nodes by moving processing load from overloaded nodes to less loaded nodes. After successful analysis of the data in the *data analysis* module, the results are shared with the user through the *visualization* component.

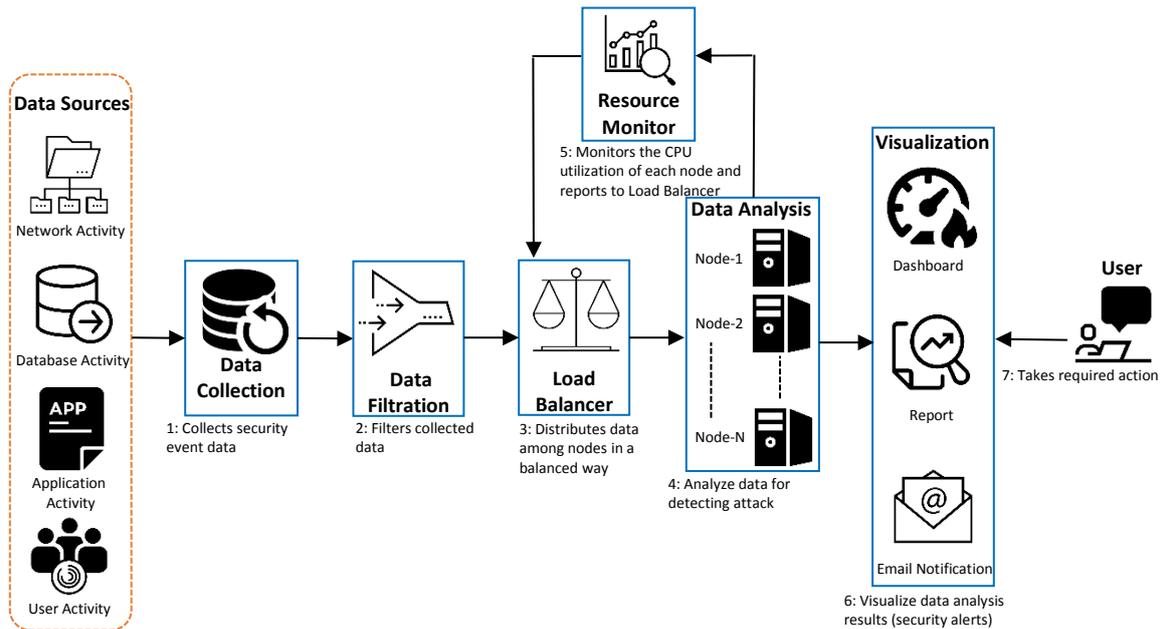

Fig. 17. Dynamic load balancing tactic

*Constraints*. This tactic assumes that a cluster has the processing capacity; otherwise, there is no chance of balancing the load if all nodes are already utilizing 100% of their CPU power. Furthermore, this tactic requires an efficient mechanism for selecting a target node and deciding the amount of data to be moved from the overloaded node to the target (under-loaded) node.

*Examples*. The following systems implement Dynamic Load Balancing tactic.

- GSLAC [S7]: During runtime, the resource monitor monitors the workload of the nodes and periodically updates the latency list that shows the CPU utilization of the nodes. When the difference between CPU utilization among the nodes crosses the threshold, the resource monitor alerts the load balancer. The load balancer uses dynamic hot spots migration [68] to rebalance the workload among the nodes.
- Cloud Bursting [S34]: This system leverages Dynamic Load Balancing tactic in a slightly different way. The resource monitor monitors the workload of the local cluster of nodes and keeps providing status information to a load balancer. Upon arrival of a new stream of security event data, the load balancer decides whether to launch the data analysis job locally or burst it to other clusters on a cloud.

*Dependencies*. In principle, the Dynamic Load Balancing tactic does not require any other tactic for its implementation. However, in a security analytic system, it will work in coordination with tactics like Unnecessary data removal (Section 5.1.2), Feature selection and extraction (Section 5.1.3), Data CutOff (Section 5.1.6), and Parallel Processing tactic (Section 5.1.4).

### 5.3.2. MapReduce

*Introduction*. This tactic has been found in all of the reviewed systems that use Hadoop platform for storing and analyzing security event data. MapReduce tactic provides a programming framework for scaling software applications across a cluster of multiple nodes. Although MapReduce is a well-documented programming framework, this tactic captures the contribution of MapReduce relevant to the enhancement of scalability. The tactic abstracts various issues of scalability that include complex parallelization, synchronization, and communication mechanisms. In addition to security, MapReduce is also a widely accepted framework in other big data analytics domains such as bioinformatics [69, 70], astronomy [71, 72], and healthcare [73].

*Motivation*. In order to scale-out a security analytic system for handling the huge amount of data, a simple solution is to add more hardware resource to a system. The additional hardware can be added to the same physical machine (vertical scaling) or it can be added as a separate physical machine (horizontal scaling). It is quite challenging to efficiently handle complex parallelization, synchronization, communications, and resource utilization with the addition of hardware resources to an existing system. Therefore, a framework is required to handle these outstanding issues.

*Description.* The main components of the MapReduce tactic are shown in Fig. 18. The *data collection* component collects the security event data from one or multiple sources. The collected data is forwarded to *data filtration* component for removing data that does not contribute to attack detection. The filtered data is partitioned by the *master nodes* to store it in the HDFS files of *data nodes*. The *mapper* inside each data node reads its assigned HDFS block of the data for processing. It is worth noting that the number of mappers does not depend upon the number of data nodes, rather, it depends upon the number of input data blocks. The mappers process data simultaneously in the form of key-value pairs (key, value) to generate intermediate results (key, values list). The intermediate results are sorted, collated, and passed to the *reducers*, which merge and aggregate the intermediate results to produce the final output. The number of mappers and reducers running simultaneously depends upon the capacity, workload, and users' recommendations. If a system experiences an increased data input, additional hardware resources (data nodes) can be easily added to a cluster to handle the increased workload. For more details on the architecture of Hadoop and MapReduce, readers can consult these souces [74, 75].

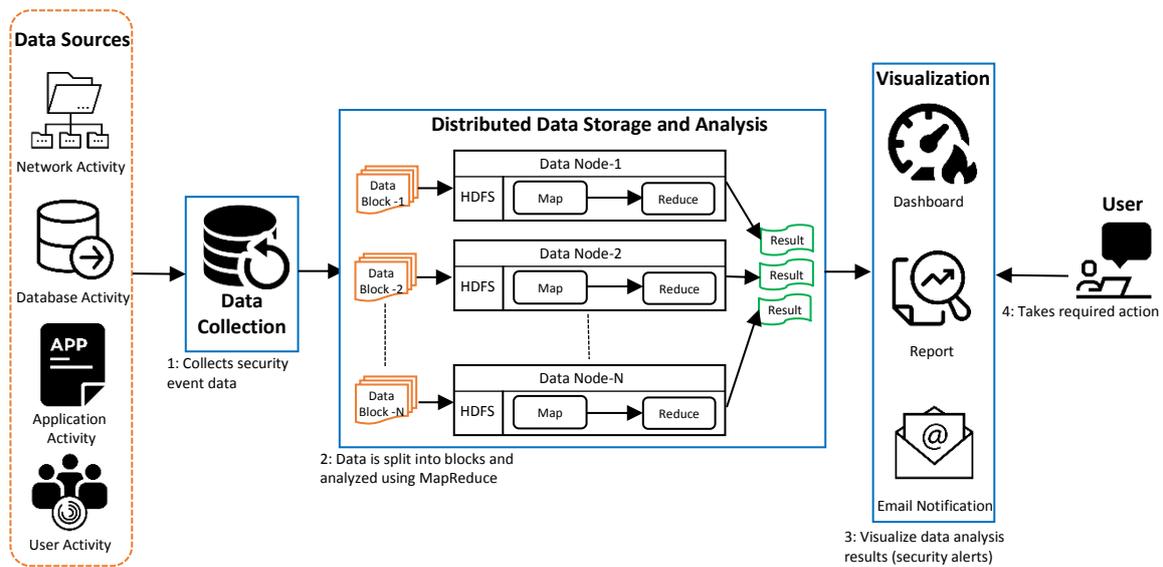

Fig. 18. MapReduce tactic

*Constraints.* This tactic assumes that additional hardware is available for horizontal scaling of a system. Furthermore, the introduction of overhead introduced due to disk read/write operations is a pressing issue with MapReduce, which can significantly prolong the response time of a security analytic system.

*Example.* MapReduce tactic is used by a number of the reviewed systems, however, we describe a few of them to illustrate its contribution in achieving scalability.

- Traffic Measurement and Analysis with Hadoop [S70]: This system uses heuristic algorithm [76] that enables mappers to read packet records from HDFS based on timestamp bit of a packet header. In order to evaluate scalability, 1 TB of security event data is analyzed by a cluster with varying number of nodes from 5 to 30. It is observed that system performance improves in proportion to the resource allocation. For example, the analysis completion time decreases from 71 minutes with five nodes to almost 36 minutes with 10 nodes.
- IDS-MRCPSO [S31]: This system uses Particle Swarm Optimization clustering algorithm [77] with MapReduce. To investigate scalability, the system is implemented with different number of nodes. It is found that the speedup time increases linearly in the start from 2 to 8 nodes, but it starts to diverge from linear while moving from 8 to 16 nodes. This diversion is attributed to Hadoop framework i.e., starting MapReduce jobs and storing intermediate results.
- Extreme Learning Machine [S24]: This security analytic system combines the linear scaling capability of Extreme Learning Machine algorithm [55] and MapReduce. The system has been implemented with 15, 20, 25, and 30 nodes to evaluate its scalability. The system shows linear scaling from 15 to 25 nodes but diverges from linear scaling after 25 nodes. The authors argue that such diversion is experienced due to the increased communication between the nodes.

*Dependencies.* As such this tactic does not require any other tactic for its implementation.

5.4. Reliability

This section reports the architectural tactics that are related to Reliability quality attribute.

**5.4.1. Data ingestion monitoring tactic**

*Introduction.* The Data Ingestion Monitoring tactic is found in Streaming-based Threat Detection System [S1], Malicious IP Detection [S18], and GPGPU-based IDS [S35]. This tactic monitors the flow of data from data collector into the computing servers in real-time data streaming systems. If the speed of data influx becomes too fast such that it can crash the computing server, this tactic automatically blocks the influx of data into the computing server.

*Motivation.* A large amount of security event data is collected from various sources in an enterprise. These sources include but not limited to network log data, user activity data, application activity data, and host access logs. This large sized data is generated at a high speed and fed directly into a security analytic system that facilitates real-time analytics. However, sometimes, data might be generated and collected at a speed that is beyond the capacity of computing cluster to process the data and would lead to crashing the computing server. Therefore, a tactic is required to monitor the speed of data generation and control influx of data into the computing servers.

*Description.* The main components of the Data Ingestion Monitoring tactic are shown in Fig. 19. The *data collector* component collects data from various sources using different tools such as Wireshark for collecting network traffic data. The data collected by various nodes of the *data collection* module is moved to the *distributed data storage and analysis* component through *data ingestion monitor*. The role of *data ingestion monitor* is to keep the real-time data streaming into the *distributed data storage and analysis* cluster. The *data ingestion monitor* monitors the speed of influx to the distributed cluster. If the speed of data influx from *data collector* becomes so high that it is beyond the capacity of the computing server and may lead to crashing the computing cluster, the monitor will block the flow of data into the computing cluster. The *data ingestion monitor* adjusts the flow of data between *data collection* component and *distributed data storage and analysis* component. The stream of security event data fed into the *distributed storage and analysis* component is analyzed for detecting cyber attacks and the results are displayed to the user through the *visualization* component.

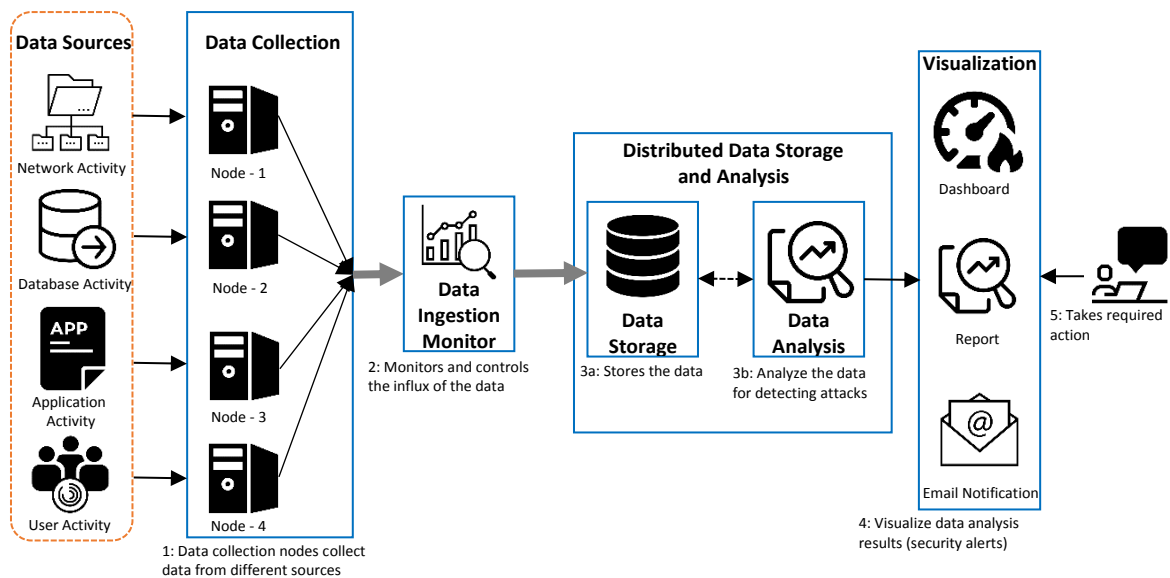

Fig. 19. Data ingestion monitoring tactic

*Constraints.* Data Ingestion Monitoring Tactic requires expertise for installation of data ingestion monitor, which is a challenging task, especially in highly distributed setup. Furthermore, the tactic also requires more investment for setting up a monitoring server. The tactic is best fit for security analytic systems that are deployed in enterprises, which generate a large volume of security event data at a high speed and stream it directly into the analytic module for real-time processing.

*Example.* The Data Ingestion Monitoring tactic is incorporated in the following systems.

- Malicious IP Detection [S18]: This system demonstrates how Data Ingestion Monitoring tactic prevents computing server from getting crashed. In this system, Flume server [79] is used as data ingestion monitor to monitor and control the influx of data into the computing cluster. The authors have evaluated the system with four data influx data rates (i.e., 0.4 million records/min, 0.7 million records/min, 0.8 million records/min, 0.85 million records/min). It has been reported that the computing server has idle capacity up to 0.8 million records/min, however, as the speed is increased to the 0.85 million records/min, the computing server reaches its maximum limit. Increasing the data influx rate beyond the 0.85 million records/min would crash the computing server. Therefore, at 0.85 million records/min, the data ingestion monitor controls the speed from increasing any further.
- Streaming-based Threat Detection System [S1] and GPGPU-based IDS [S35]: Both these systems implement Data Ingestion Monitoring tactic through Flume service, however, these studies have not evaluated the impact of the tactic on the reliability of the system.

*Dependencies.* The Data Ingestion Monitoring tactic does not depend upon any other tactic; however, it can better be consolidated when implemented together with Secure Data Transmission tactic (Section 5.5.1), which will help get the data to be monitored in its original form.

### 5.4.2. Maintaining multiple copies

*Introduction.* In principle, the Maintaining Multiple copies tactic can be found in any system that leverages Hadoop framework, however, only seven [S2], [S6], [S10], [S13], [S17], [S31], [S71] of the reviewed papers explicitly demonstrate how this tactic can make a system more reliable. This tactic enables Hadoop's HDFS to maintain multiple copies of each data block. These copies are placed on different nodes in a Hadoop infrastructure. In case of a node failure, the computation is diverted to another node that hosts the copy of the data block [78].

*Motivation.* During data processing, a Hadoop node may go down or fail due to several reasons such as RAM crash, power shutdown, and hard-disk failure. In such a case, data on the same node will no longer be accessible and a user has to wait until the issue is addressed. Therefore, a mechanism is required that can enable a security analytic system to have access to the data even if a node fails.

*Description.* The main components of the Maintaining Multiple Copies tactic are shown in Fig. 20 with the numbers showing the sequence of operations. The *data collector* component collects security event data from different sources, which is stored in files. The files are divided into blocks of same sizes. The default block size is 64 MB; however, the block size can be customized. For the sake of understanding, only one file split into four blocks is shown in Fig. 20. After dividing files into blocks, the blocks are stored on different nodes (i.e., machines). By default, three copies of each data block are created. However, the number of replicas created can be configured. Each data block is stored on three different machines. For example, Block 1 is stored on Node 1, Node 4, and Node 6. Then the *data analysis* component reads the data blocks for security analysis. If a data block is not accessible at any instance (for instance block 1 on Node 1), the same data block will be accessed from another node (for instance block 1 from Node 4). Once the data is analyzed in a reliable manner, the results are shared with the user through the *visualization* component.

*Constraints.* With the incorporation of the Maintaining Multiple Copies tactic, the cost of storing the same data increases. This tactic may also introduce some data inconsistencies.

*Example.* Following are a couple of examples that illustrate the implementation of the Maintaining Multiple Copies tactic.

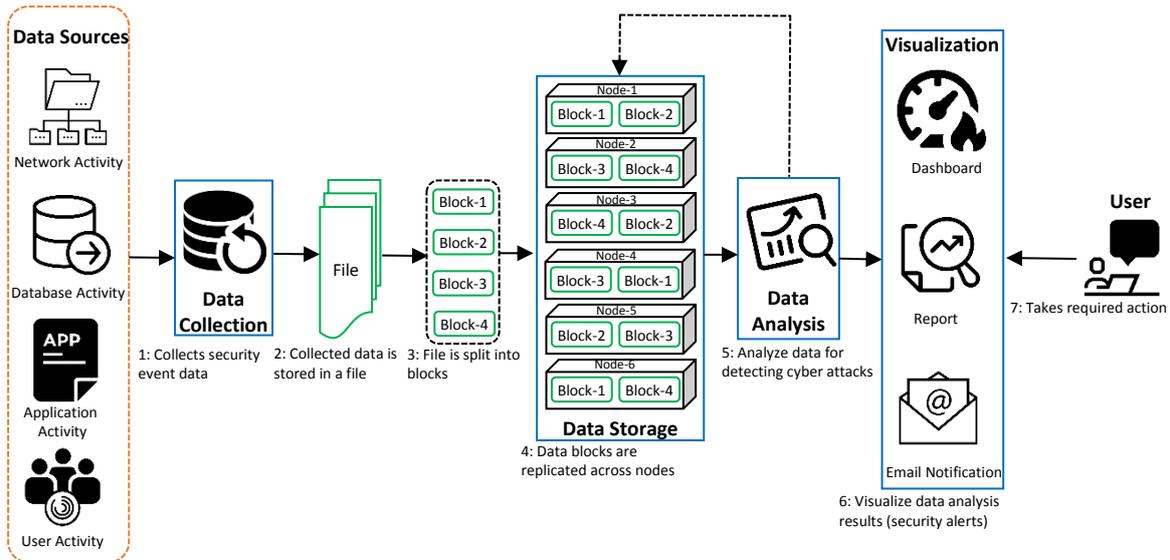

Fig. 20. Maintaining multiple copies tactic

- Honeypot-based Phishing Detection [S17]: In this study, the impact of replication factor (i.e., number of copies) on the performance of a security analytic system is investigated. The system is tested with one, two, and three replicas. It is observed that a system's response time is best with three replicas (21 min), followed by two replicas (22 min). With one replica, a system can have the lowest response time (25 min) and no reliability in case of a node failure. The difference in response time is due to the number of non-local accesses incurred in each case, which is 4 with three replicas, 7 with two replicas, and 113 with one replica.
- Reliable Traffic Analysis [S71]: This system is tested with two node failure scenarios – one where node executing map task fails and another where node executing reduce task fails. The node running map task is forced to reboot while the node running reduce task is shutdown. The map and reduce tasks are migrated to other nodes having the replicas of the data. It has been observed that the system remained available, however, it takes little more time in completing the tasks. For example, for flow count of 3.2 million, the completion time without nodes failure is 220.2 seconds while with node failure the completion time is 380.2 seconds.

*Dependencies.* The Maintaining Multiple Copies tactic does not require any other tactic for its implementation, however, it can better be consolidated when implemented together with Secure Data Transmission tactic (Section 5.5.1), which will help get the data to be replicated in its original form.

### 5.4.3. Dropped NetFlow detection

*Introduction.* Dropped NetFlow Detection tactic has been found in Hybrid Stream and Batch Analyzer [S54]. NetFlow is a sequence of packets that typically share seven values, which include source IP, destination IP, source port, destination port, IP protocol, type of service, and ingress interface [38]. This tactic helps security analytic systems, which rely on collecting and analyzing NetFlow data for detecting attacks, to detect if the data collector has missed some NetFlows. The tactic monitors the NetFlow Sequence Numbers [80] and if they are found out of order, a warning message specific to that stream is logged. This tactic is important for detecting faults that can have severe consequences for reliable data collection.

*Motivation.* For security analytic system relying on NetFlow data, it is important to collect and analyze each NetFlow. Missing a single NetFlow may result in missing the detection of an attack (e.g., a single flow record may summarize malicious transfer of a large amount of data). In the existing network dynamics, it is quite possible that some NetFlows are missed due to several reasons that include (a) network dropping packets; (b) router not able to handle the volume of traffic; and (c) data collector of a security analytic system not able to handle the volume of traffic. Hence, it is quite crucial to monitor and detect whether a security analytic system is analyzing all NetFlow or there are some NetFlows going to missing. In the latter case, a security administration needs to trace the source causing loss and fix the issue.

*Description.* The main components of the Dropped NetFlow Detection tactic are shown in Fig. 21 along with the number indicating the sequence of the operations. The network traffic is passing through the router shown in the figure. A *NetFlow collector* is connected to the router, which collects the NetFlows and stores them in the *NetFlow storage*. During the NetFlow collection process, *NetFlow sequence monitor* component is monitoring the sequence numbers embedded (by design) in the NetFlows. If the sequence numbers are found out of order at any stage, the *NetFlow sequence monitor* generates a *warning message* indicating the missing flow in the particular stream of NetFlow. The warning message is logged alongside the specific stream in the *NetFlow storage* to specify that the stream of NetFlow has some flows missing that might be crucial in relation to detecting an attack. At the same time, a warning is displayed to a security administrator through the *visualization* component. Then a security administrator may take the required actions for fixing the issue due to which some NetFlows may get dropped.

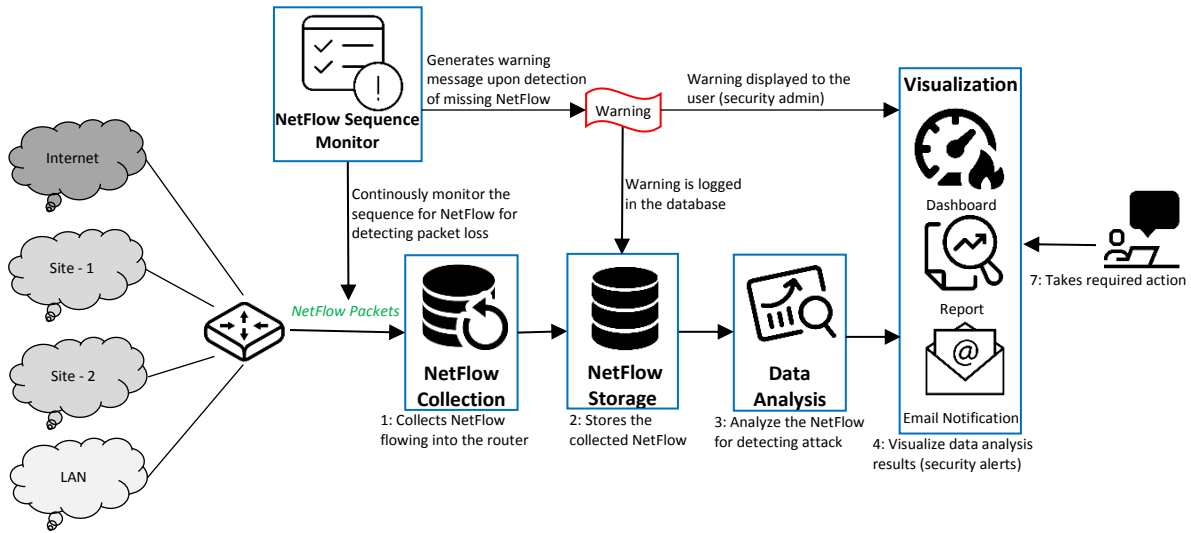

Fig. 21. Dropped NetFlow Detection Tactic

*Constraints.* The NetFlow sequence monitoring should cope with the speed at which NetFlow is collected so that additional delay for monitoring can be avoided. This tactic is best fit for security analytic systems that monitor highly busy networks where there is a chance of packet loss. It is worth noting that this tactic only detects the loss of NetFlows. For mitigation, i.e., identifying the source of loss and fixing, it requires a separate mechanism.

*Example.* Hybrid Stream and Batch Analyzer [S54] is the example systems that have implemented the Dropped NetFlow Detection tactic. The system has multiple NetFlow collectors that run on the probe nodes. Each NetFlow collector has two associated AMPQ queues [81] – one for the data stream and another for log information. The system experiences loss of NetFlows due to link aggregation [82]. In order to detect the missing NetFlows, the sequence numbers of the NetFlows are monitored and when they are found out of sequence, a warning message specific to the stream is placed in the log information queue.

*Dependencies.* This tactic can be implemented independent of any other tactics.

5.5. Security

This section reports the architectural tactic that is related to Security quality attribute.

**5.5.1. Secure data transmission**

*Introduction.* The Secure Data Transmission tactic has been reported in Cloud-based Threat Detector [S10] and PKI-based DIDS [S73]. This tactic ensures secure transmission of security event data from data collection nodes to the data analysis nodes.

*Motivation.* A security analytic system protects a critical infrastructure from cyber attacks, however, it in itself also requires some security measures. In most cases, the data collection and analysis operations reside on separate nodes (physical machines). The nodes hosting data collection operation are placed in different regions within an organization [83]. The data collected by these nodes is transmitted to the nodes (cloud cluster) hosting the data analysis operation. It is quite possible that an attacker intercepts this data transmission (e.g., with a sniffing tool) to tamper with or spoof the data [84]. The consequences of such interception can be catastrophic if an attacker manages to tamper with the data critical for detecting an attack. Therefore, it is important to guarantee secure transmission of data from data collection nodes to the cloud cluster responsible for data analysis.

*Description.* Fig. 22 shows the main elements of the Secure Data Transmission tactic. The nodes for collecting security event data are placed in different regions for collecting different types of data. Some collect network traffic, others collect database access information and so on. Security measures are applied to the collected data to ensure its secure transfer from *data collection* component to the *data storage and analysis* component. The security measures incorporated vary from system to system. Some systems prefer to encrypt the collected data and then transfer the data in encrypted form. Other systems use Public Key Infrastructure (PKI) to ensure secure exchange of data and verification of the party sending the data. As data is received by the *data storage and analysis* component in a secure way, the data analytic operations are applied to analyze the data for detecting attacks. The results of the analysis are presented to users through the *visualization* component.

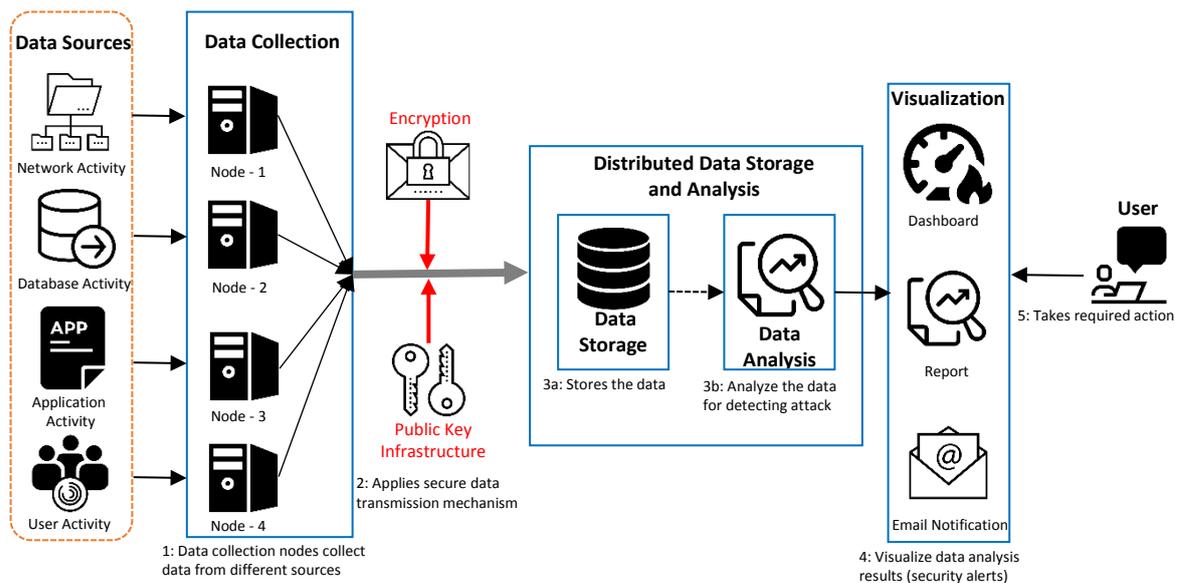

Fig. 22. Secure data transmission tactic

*Constraints.* The introduction of security measures in communication among different components will slow down the data transmission process and eventually the response to an attack. It is also worth noting that the Secure Data Transmission tactic protects data only in motion state, which means that additional measures will be required to secure the data in use and in rest state both at the collection nodes and the storage and analysis nodes. For instance, the injection of false data in one of the data collection nodes can have disastrous consequences in terms of attack detection.

*Example.* The two systems that have implemented the Secure Data Transmission tactic incorporate some kind of security measure for secure transmission of data from data collection component to the data storage and analysis component.

- Cloud-based Threat Detector [S10]: In this system, all communications between various modules take place in encrypted form, which is achieved by adopting HTTPS protocol. In order to ensure the authenticity of access, session tokens are checked and verified.
- PKI-based DIDS [S73]: This system deploys Public Key Infrastructure [85] between data collection nodes and data storage and analysis nodes. According to this security measure, digital certificates are used to encrypt data to be transmitted from data collection to data storage and analysis. It also enables the data collection and data storage component to verify the authenticity of data collection nodes. This way, even if an attacker intercepts the transmission, he will not be able to tamper with the confidentiality or integrity of the data.

*Dependencies.* This tactic can be implemented independent of other tactics.

## 5.6. Usability

This section reports the architectural tactic that is related to Usability quality attribute.

### 5.6.1. Alert ranking

*Introduction.* Alert Ranking tactic has been reported in Hunting attacks in the dark [S41]. This tactic ranks the alerts generated by a security analytic system based on the dangerousness of the alert. This ranking improves the usability of a system by enabling users (e.g., security administrator) to respond to the alerts on a priority basis.

*Motivation.* A security analytic system can generate a large number of alerts [86, 87]. In around 90% of the cases, the alerts are either false positives or of low importance [88]. It is also well understood that the more dangerous an alert is the more challenging it is to respond and mitigate its effects [S41]. If a security administrator responds to alerts in a sequence in which the alerts are generated, it is quite likely that serious alerts requiring an immediate response will be responded quite late. On the other hand, if a security administrator himself assesses the seriousness of alerts, it will be quite time-consuming for an administrator. The late response because of the large number of alerts will be quite detrimental in the sense that it will be more challenging to mitigate the effects of an attack. Therefore, it is important to incorporate a mechanism in a security analytic system that can help the security administrator to prioritize the response to the more serious alerts.

*Description.* The main components of the Alert Ranking tactic are shown in Fig. 23 with the numbers indicating the sequence of the operations. The *data collection* component collects security event data from different sources, which is pre-processed by the *data pre-processing* component. The pre-processed security event data is forwarded to the *data analysis* module, which analyzes the data for detecting cyber attacks. The results of the analysis (i.e., alerts) are forwarded to the *alert ranking* component, which ranks the alerts based on a predefined criterion to assess the impact of the alert on the overall organization's infrastructure. The criterion for ranking the alerts depends upon an organization. For example, the ranking criteria for an organization vulnerable to DoS attacks will be different from an organization vulnerable to brute force attacks. Finally, the ranked list of simple and easy-to-interpret alerts is shared with security administrators through the *visualization* component, which eases the job of a security administrator to first respond to the alerts on top of the rank list as these alerts are anticipated to be more dangerous and consequential.

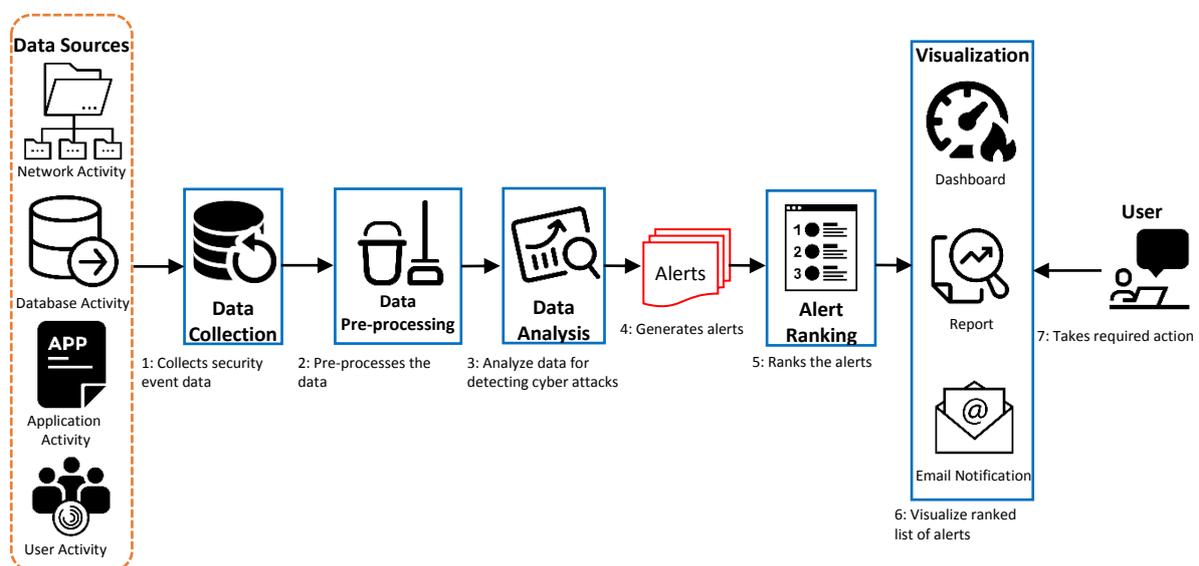

Fig. 23. Alert ranking tactic.

*Constraints.* The Alert Ranking tactic requires that the functionality responsible for ranking the alerts should not be computationally expensive, otherwise it will cancel the benefit of quick response acquired through prioritized response to dangerous alerts. Moreover, all ranking algorithms are not equally accurate [89]. Therefore, it is important to carefully select the ranking algorithm (criteria) that can accurately rank the alerts based on the specific security requirements of an organization.

*Example.* Hunting attacks in the dark [S41] is an example system that has implemented the Alert Ranking tactic. This system ranks the generated alerts based on the amount of traffic that belongs to the alert. The underlying consideration for this criterion is that alerts indicating a sudden increase in the network traffic (e.g., flooding-based attacks) are the most serious ones. The system uses the number of packets and the number of bytes belonging to each alert to calculate Dangerousness Index (DI) for each alert using the formula (i.e., DI(alert) = Number of packets + Number of bytes). The bigger the DI of an alert is, the more dangerous is the alert. Based on the DI scores, the alerts are ranked and presented to the security administrator.

*Dependencies.* The Alert Ranking tactic does not entirely depend on any other tactic, however, this tactic is usually implemented together with tactics that can support the data analysis component such as Parallel Processing tactic (Section 5.1.4) and Attack Detection Algorithm Selection tactic (Section 5.2.3).

## 6. Discussion

Section 3, 4, and 5 have presented the main findings of our review. In this section, we discuss the main findings to provide a few recommendations for utilizing the identified tactics for designing and implementing big data cybersecurity analytics systems and highlight some gaps for the future research in this area.

6.1. Mapping of tactics to the components of security analytics systems

Fig. 24 shows the mapping of the architectural tactics reported in Section 5 onto the major components of a security analytic system: data collection, data storage, data pre-processing, data analysis, and visualization. The objective of this mapping is twofold.

First - it is expected to help a reader to understand which component(s) or stage(s) are highly critical from an architectural point of view. For example, most of the architectural tactics are focussed on the data analysis component; it means that a software architect must invest a significant amount of time and energy in designing this component. Similarly, the mapping reveals that the design of the visualization component has received the least amount of attention in the systems reported in the reviewed papers for this study. It is also apparent from the figure that the security tactic (Section 5.5.1) is applied between the data collection and data analysis components. However, we support the view of Anderson [90] that security will be equally critical for data transmission between any two components in a distributed system.

Second - this mapping provides a relationship between the quality attributes and different components. This type of visual information is expected to help a reader understand which components are important for achieving certain quality attributes. It is evident from the figure that the Accuracy related tactics (i.e., Section 5.2) are completely focussed on the data analysis component. On the other hand, the Reliability related tactics (Section 5.4) are associated with multiple components such as data collection, data storage, and data analysis.

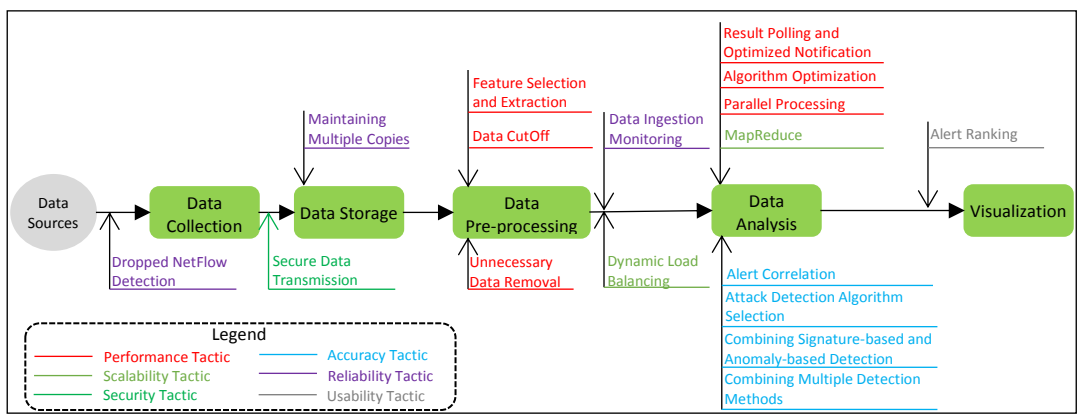

Fig. 24. Mapping of tactics to the components of big data cybersecurity analytic system

6.2. Under-addressed quality attributes

Section 4 reports our analysis of the importance of specific quality attributes for security analytic systems. Apart from Performance, Accuracy, Scalability, Reliability, Security, and Usability, our analysis of the reviewed papers has also revealed that quality attributes such as interoperability, adaptability, modifiability, generality, stealthiness and privacy assurance are also critical for security analytic systems. However, we could not find explicitly reported architectural support for achieving these qualities in the reviewed papers. It is worth mentioning that some of the reviewed papers briefly mention the need of achieving interoperability, modifiability, adaptability, generality and privacy assurance for security analytics systems. For example, the papers [S19] and [S42] advocate the use of open source tools for achieving interoperability. The papers [S10] and [S29] recommend higher modularity for achieving interoperability and modifiability. For generality, the papers [S12] and [S23] encourage the incorporation of diverse data sources and frequently updating attack patterns. However, the reported recommendations are quite generic and abstract. Therefore, we assert that there is an important need for research and development efforts aimed at providing architectural support for interoperability, modifiability, adaptability, generality, stealthiness and privacy assurance in security analytics systems.

6.3. Tactics evaluation

It is equally important to assess the impact of the elicited architectural tactics on a security analytic system. Since the tactics have been elicited from specific implementations (e.g., security analytic system detecting DoS attack) using different datasets, their incorporation and potential impact cannot be generalized to all types of security analytics systems. It is also worth noting that the primary studies from which the tactics have been extracted do include an evaluation phase, however, it does not report the evaluation of each of the identified architectural tactic. For example, the impact of Combining Multiple Detection methods tactic (Section 5.2.4), Alert Correlation tactic (Section 5.2.1), Dynamic Load Balancing tactic (Section 5.3.1), Dropped NetFlow Detection tactic (Section 5.4.3), Secure Data Transmission tactic (Section 5.5.1), and Alert Ranking tactic (Section 5.6.1) have not been evaluated. For those researchers who may be interested in conducting empirical assessment of the potential impact of the identified tactics on a security analytics system, we can propose options. For example, a survey of the expert software architects can be conducted to find out their experience-based assessment of the correctness and the potential impact of the identified tactics. Another option can be to set up a series of empirical experiments, using the same dataset, aimed at quantitatively evaluating the potential impact of the identified tactics.

6.4. Quality trade-offs among tactics

It is well known that architectural tactics may positively support one set of quality attributes but may have negative impact on another set of quality attributes [91]. For example, the Alert Correlation tactic (Section 5.2.1) helps achieve Accuracy quality attribute but can have negative impact on Performance quality attribute. Similarly, the Maintaining Multiple Copies tactic (Section 5.4.2) improves Reliability but increases the overall storage cost. Moreover, an architectural tactic can have a positive impact on multiple quality attributes unlike its association with one quality attribute as reported in this review. For instance, the Parallel Processing tactic (Section 5.1.4) improves both Performance and Reliability of a security analytic system. Understanding the trade-offs among tactics and undertaking decisions that make suitable trade-offs is a challenging task for an architect [92]. That is why we encourage more investigation for assessing the impact of the reported architectural tactics on various quality attributes of a security analytic system. Such an assessment will be useful for an architect to decide upon the application of the tactics according to the specific functional and non-functional requirements.

6.5. Dependencies among tactics

In addition to the impact of the architectural tactics on quality attributes and subsequent trade-offs, it is important to reflect upon the dependencies among the architectural tactics. The reason for such exploration is that the architectural tactics cannot be applied in isolation and may require other tactics to be implemented. As mentioned in Section 5, architectural tactics may depend on other tactics for achieving their respective objectives. For example, the Feature Selection and Extraction tactic (Section 5.1.3) depends upon the Parallel Processing tactic (Section 5.1.4) for speeding up the process of feature selection and extraction. Furthermore, a combination of tactics may complement each other in a way that allows an architect to drop one tactic in favour of another tactic to resolve a conflict between two quality attributes. For example, the Combining Multiple Detection Methods tactic (Section 5.2.4) and the Combining Signature-based and Anomaly-based Detection tactic (Section 5.2.2) complement each other to achieve Accuracy quality attribute, however, the incorporation of both will increase the response time. Therefore, an architect may consider dropping either of them depending on a system's requirements. Whilst the dependencies among tactics have been reported to a certain level in Section 5, this review has enabled us to assert that there is likely to be much deeper dependencies among the identified architectural tactics; another area for empirical research on architectural tactics for security analytics systems.

6.6. Academic-Industry disconnect

Our study has helped us to determine a general lack of engagements between academia and industry for the work reported in the reviewed papers. There are only seven studies ([S18], [S19], [S20], [S23], [S36], [S49], [S59]) that have shown some engagements in terms of data sharing and practical deployment of security analytic systems in an industrial setting. We recommend that the academic-industry engagements be more intense and rigorous for mutual benefits of devising and evaluating innovative and new architectural tactics and principles for security analytics systems. Our recommendation has been motivated from the findings of two studies [93, 94], which emphasise that deep engagements between academia and industry is fruitful both for academia and industry. The dissemination of academic findings to industry helps practitioners to have access to findings based on systematic and rigorous work rather than relying on personal observation. At the same time, practitioners' knowledge and industrial settings can help academics to develop hypotheses and design and conduct effective experimental evaluations. We recommend that there be more and intense academic-industry engagements through large scale empirical and field studies using multi-methods approaches (such as case studies, action research, interviews, and surveys) that are expected to help and disseminate evidence-based findings to support security analytics systems.

6.7. Comparison among big data processing frameworks

As evident from Fig. 4a, 67% of the security analytics systems reported in the reviewed papers leverage mainly Hadoop followed by Spark (19%) and Storm (14%). It is also worth noting that only six papers ([S17], [S41], [S54], [S59], [S62], [S66]) report some comparison of these frameworks when used for security analytics systems. However, the comparison is either between Hadoop and Spark ([S17], [S41], [S62]) or between Hadoop and Storm ([S54], [S59], [S66]). We assert that there is an important need of more extensive comparison among all three frameworks in the context of security analytics systems. Such a comparison will help practitioners to gain valuable information for selecting the most suitable framework for their security analytic systems. Moreover, it would be interesting to explore the applicability of Spark and Storm for the systems (i.e., 67% of cases in the reviewed paper) where Hadoop has been used. Fig. 4b indicates an increasing interest in leveraging Spark and Storm. This interest should be carried further to rigorously investigate the use of Spark and Storm for security analytics systems.

## 7. Related work

The research and practice for design and developing security analytics systems have been drawing an increasing amount of attention of researchers and practitioners. There have been a few reviews of the literature on security analytics from different perspectives. However, we have not come across a literature review on architectural aspects (i.e., quality attributes and architectural tactics) of security analytic systems; nor has there been any review conducted using Systematic Literature Review methodology. We briefly summarize some of the existing reviews on the topic of security analytics in order to highlight the positioning of our work. Kaj et al. [95] present a survey on analytics for cybersecurity, which is focussed on the techniques adopted for cybersecurity analytics. The authors have proposed a taxonomy that categorizes the underlying techniques into descriptive, diagnostic, visual, predictive, and prescriptive techniques. Prescriptive techniques are highlighted as being of high potential and advocate for its exploration. Lidong and Randy [96] provide an overview of various aspects of big data analytics for network intrusion detection. Their review concisely covers the big data processing frameworks (e.g., Spark and Storm), data pre-processing, and machine learning for security analytics and motivates the incorporation of big data tools and technologies for cybersecurity analytics. Tariq and Uzma [97] have conducted a review that emphasizes the imminent application of big data analytics for cybersecurity. Their review sheds light on different types of cyber attacks (e.g., botnets and phishing), various sources from which security event data can be collected, sample outputs of a security analytic system, and some commercially available security analytic solutions. Jeong et al. [98] have investigated various challenges and potential solutions for IDS that leverage the Hadoop processing framework. The reported challenges include emerging attack patterns, high implementation cost, and the large volume, high velocity and heterogeneous nature of security event data. The authors argue that the use of Hadoop, MapReduce and NoSQL databases, and incorporation of open source software will help address the aforementioned challenges. Rasim and Yadigar [99] have reviewed the state-of-the-art of big data analytics for cybersecurity for reporting big data tools and technologies that can be leveraged in security analytics. Their review highlights various security analytics challenges such as privacy, APT detection, cryptography, huge amount of dataset, data provenance, security visualization, and lack of skilled personnel.

The work that is most closely related to our work is that of Ricard et al. [100], which reports the major challenges faced by IDS for analyzing a large amount of heterogeneous data. The authors highlight that complex machine learning, lack of evaluation datasets, and feature selection are significant challenges for intrusion detection. Furthermore, the review recommends the correlation of security events, the collection of data from diverse sources, and incorporation of big data tools and technologies for efficient intrusion detection. However, this work is neither architecture-centric nor does it follow a systematic process. To the best of our knowledge, our work is the first SLR focussed on the architecture and architecturally significant requirements (quality attributes) of security analytic systems. We believe that our work is the first of its kind in terms of systematically determining and rigorously developing a catalogue of architectural tactics that is expected to provide significant insights and utility for designing and evolving security analytic systems that deal with huge amount of data.

## 8. Limitations

Whilst we have designed and conducted this SLR by following the guidelines of [14], there are certain limitations that need to be considered while examining its findings.

Like any other SLR, our SLR also faced the issue of primary studies being missed. To minimize the potential impact of this limitation, we searched six digital repositories for selecting primary studies. According to [14] and [101], these six repositories are the most relevant to literature on computer science and software engineering.. We did not impose any restrictions on the publication date of the primary studies. To ensure the reliability of the search string, we tested the search string through pilot searches and iteratively improved it until the searches returned the papers that included the papers that we already knew to be relevant. Then we also employed the snowballing technique [16] that enabled us to review the references of the primary studies obtained via the search to identify further relevant studies.

The selection of the papers from the pool of papers retrieved from the digital databases can be influenced by the subjective judgement of researchers of an SLR. Whilst we made the inclusion and exclusion criteria (reported in Section 2.3) as explicit as possible, the lack of well-defined terminologies on the architecture-centric approach in the studies made the selection quite challenging. To address this issue, the reasons for inclusion and exclusion of the papers were recorded and reviewed by the authors. We also tried to reduce the potential misinterpretations to a minimum by following a multi steps selection process (i.e., title-based, abstract-based, and full text based).

A researcher's bias in data extraction is another limitation faced by SLRs that influences their findings. To minimize the potential impact of this threat, we designed a data extraction form (Table 4) for consistently extracting the data from reviewed papers. As mentioned in Section 2, we employed quantitative and qualitative analysis techniques for producing the results reported in this SLR. The possibility of interpretation bias cannot completely be ruled out in such analysis as sometimes the extracted data item is not obvious (e.g., constraints and dependencies associated with tactics), and some level of interpretation is required. To mitigate the researchers' bias in data interpretation, wherever possible we explored the antecedent and subsequent publications of the studies, websites of the tools developed based on the study, and blogs of the authors of the reviewed papers.

## 9. Conclusion

Motivated by the growing significance of big data analytic systems for cyber security, it is important to systematically gather and rigorously analyze and synthesize the literature on architectural strategies used for designing such systems. Given there was a general lack of an overview of architectural strategies for big data cybersecurity analytic systems, we have conducted a systematic literature review of big data cybersecurity analytic systems from an architectural perspective. Based on the review of 74 relevant papers, we have identified and explained 12 quality attributes as critical for a big data cybersecurity analytic system. Furthermore, we have also identified and codified 17 architectural tactics for supporting the required qualities (i.e., performance, accuracy, scalability, reliability, security, and usability) in a big data cybersecurity analytic system.

The results of this SLR can have several implications both for researchers and practitioners. For researchers, our SLR have identified a number of areas for future research. The demographic findings are of potential value for researchers to shape their future research directions. For example, it is quite evident that the incorporation of big data tools and technologies in security analytics is observing an upward growth. Similarly, along the course of years, spark is getting more popular as compared to Hadoop. Given that important quality attributes such as interoperability, modifiability, adaptability, generality, stealthiness and privacy assurance lack architectural support, we assert that researchers need to explore various options for developing such an architectural support. Other areas for potential research include empirical evaluation of the reported architectural tactics, trade-off and dependency analysis among the tactics, and comparative analysis among big data processing frameworks (e.g., Hadoop, Spark, and Storm) when employed in a big data cybersecurity analytic system.

For practitioners, the identification of the most relevant quality attributes and a catalogue of architectural tactics can be used as a reference in designing big data cybersecurity analytic systems. Given that the elicitation of the non-functional requirements (e.g., privacy, adaptability, scalability etc.) is a challenging task, practitioners can benefit from our reporting of important quality attributes supported by qualitative and quantitative reasoning to establish non-functional requirements for their big data cybersecurity analytics systems. It is worth mentioning that this SLR is a first step towards guiding software architects and software engineers to efficiently architect security analytic systems. For concretizing the body of knowledge and materializing practitioners' trust through empirical evidence, we plan to establish an experimental setup of distributed computing complemented by big data tools and technologies for rigorous empirical evaluation of the codified architectural tactics. The experimental design will aim to investigate the trade-offs and dependencies among the tactics. We are also enthusiastic to invest our future efforts in developing automation support for instantiating these tactics, which will reduce the developers' effort and speedup the development process. Along the same lines of automation, we intend to investigate the refactoring of existing security analytic systems to support the incorporation of our tactics in the system. We also aim to further facilitate practitioners by developing a body of knowledge on strengths and weakness of big data tools and technologies employed in security analytics. Such a body of knowledge will help practitioners to select tools and technologies according to their specific requirements.

# Appendix A. Primary studies selected for this SLR.

Table 7. List of included primary studies

| ID | System name | Title | Author(s) | Publication venue | Year |
|---|---|---|---|---|---|
| S1 | Streaming-based Threat Detection | A Streaming-Based Network Monitoring and Threat Detection System | Zhijiang Chen, Hanlin Zhang, William G. Hatcher, James Nguyen, Wei Yu | Software Engineering Research, Management and Applications (SERA) | 2016 |
| S2 | IP Spoofing typed DDoS detection | A Hadoop based analysis and detection model for IP Spoofing typed DDoS attack | Jian Zhang, Pin Liu, Jianbiao He, Yawei Zhang | IEEE TrustCom/BigDataSE/ISPA | 2016 |
| S3 | Stochastic Self-similarity | Detecting Anomaly Teletraffic Using Stochastic Self-Similarity Based on Hadoop | JongSuk R. Lee, Sang-Kug Ye and Hae-Duck J. Jeong | Conference on Network-based Information Systems | 2013 |
| S4 | Combining IDS Datasets | Combining intrusion detection datasets using MapReduce | Mondher Essid, Farah Jemili | Conference on Systems, Man, and Cybernetics | 2016 |
| S5 | K-means based Anomaly Detector | Anomaly Detection in Network Traffic using K-mean clustering | R. Kumari, Sheetanshu, M. K. Singh, R. Jha, N.K. Singh | Conference on Recent Advances in Information Technology | 2016 |
| S6 | Multistage Alert Correlator | Distributed Multistage Alert Correlation Architecture based on Hadoop | James Rees | International Carnahan Conference on Security Technology | 2015 |
| S7 | GSLAC | GSLAC: A General Scalable and Low-overhead Alert Correlation Method | Li Cheng, Yijie Wang, Xingkong Ma and Yongjun Wang | IEEE TrustCom/BigDataSE/ISPA | 2016 |
| S8 | Website IDS using Hadoop | Hadoop-based System Design for Website Intrusion Detection and Analysis | Xiaoming Zhang, Guang Wang | Conference on Smart City/SocialCom/SustainCom | 2015 |
| S9 | Spark-based IDS Framework | A Framework for Fast and Efficient Cyber Security Network Intrusion Detection using Apache Spark | Govind P Guptaa, Manish Kulariyaa | Conference on Advances in Computing & Communications | 2016 |
| S10 | Cloud-based Threat Detection | A Cloud Computing Based Network Monitoring and Threat Detection System for Critical Infrastructures | Zhijiang Chen, Guobin Xu, Vivek Mahalingam, Linqiang Ge, James Nguyen, Wei Yu, Chao Lu | Big Data Research | 2015 |
| S11 | Spatiotemporal Correlator | A Cloud Computing Based Architecture for Cyber Security Situation Awareness | Wei Yu, Guobin Xu, Zhijiang Chen, and Paul Moulema | Workshop on Security and Privacy in cloud computing | 2013 |
| S12 | Quasi Real-time IDS | Big Data Analytics framework for Peer-to-Peer Botnet detection using Random Forests | Kamaldeep Singh, Sharath Chandra Guntuku, Abhishek Thakur a, Chittaranjan Hota a | Journal of Information Sciences | 2014 |
| S13 | Spark-based Network Data Analysis | Network Data Analysis Using Spark | K.V. Swetha, Shiju Sathyadevan, and P. Bilna | Software Engineering in Intelligent Systems | 2015 |
| S14 | Ultra-High-Speed IDS | Real time intrusion detection system for ultra-high-speed big data environments | M. Mazhar Rathore1 · Awais Ahmad1 · Anand Paul1 | Journal of supercomputing | 2016 |
| S15 | METRIS | METIS: A Two-Tier Intrusion Detection System for Advanced Metering Infrastructures | Vincenzo Gulisano, Magnus Almgren, and Marina Papatriantafilou | Conference on Security and Privacy in Communication Systems | 2014 |
| S16 | Big Data Security Monitoring | A Big Data Architecture for Large Scale Security Monitoring | Samuel Marchal, Xiuyan Jian, Radu State, Thomas Enge | International Congress on Big Data | 2014 |
| S17 | Honeypot-based Phishing Detection | A Big Data architecture for security data and its application to phishing characterization | Pedro H. B. Las-Casas, Vinicius Santos Dias, Wagner Meira Jr. and Dorgival Guedes | Conference on BigDataSecurity-HPSC-IDS | 2016 |
| S18 | Malicious IP Detection | A Real-time Anomalies Detection System based on Streaming Technology | Yutan Du, Jun Liu, Fang Liu, Luying Chen | Conference on Intelligent Human-Machine Systems and Cybernetics | 2014 |
| S19 | AGILIS | Distributed Attack Detection Using Agilis | Leonardo Aniello, Roberto Baldoni, Gregory Chockler, Gennady Laventman, Giorgia Lodi, and Ymir Vigfusson | Journal of Collaborative Financial Infrastructure Protection | 2012 |
| S20 | Detecting DDoS in Cloud Service | A Spark-Based DDoS Attack Detection Model in Cloud Services | Jian Zhang, Yawei Zhang, Pin Liu(B), and Jianbiao He | Conference on Information security practice and experience | 2016 |
| S21 | Compression model for IDS | Efficient classification using parallel and scalable compressed model and its application on intrusion detection | Tieming Chen, Xu Zhang a, Shichao Jin, Okhee Kim | Journal of Expert Systems with Applications | 2014 |
| S22 | BotFinder | An Automated Bot Detection System through Honeypots for Large-Scale | Fatih Haltaş, Abdulkadir Poşul, Erkam Uzun, Bakır Emre | Conference on Cyber Conflict | 2014 |
| S23 | Botnets with Graph Theory | Big Data Behavioral Analytics Meet Graph Theory: On Effective Botnet Takedowns | Elias Bou-Harb, Mourad Debbabi, and Chadi Assi | IEEE Network | 2017 |
| S24 | Extreme Learning Machine | Using Extreme Learning Machine for Intrusion Detection in a Big Data Environment | Junlong Xiang, Magnus Westerlund, Dušan Sovilj, Göran Pulkkis | Workshop on Artificial Intelligent and Security | 2014 |

| ID | Name | Title | Authors | Venue | Year |
|---|---|---|---|---|---|
| S25 | Intersection-based Pattern Matching | Privacy-Preserving Scanning of Big Content for Sensitive Data Exposure with MapReduce | Fang Liu, Xiaokui Shu, Danfeng (Daphne) Yao and Ali R. Butt | Conference on Data and Application Security and Privacy | 2014 |
| S26 | WEKA-based Anomaly Detector | Anomaly detection model based on Hadoop platform and Weka interface | Baojiang Cui, Shanshan He | Conference on Innovative Mobile and Internet Services in Ubiquitous Computing | 2016 |
| S27 | Improved K-means IDS | An Intrusion Detection System Based on Hadoop | Zhiguo Shi, Jianwei An | UIC-ATC-ScalCom-CBDCom-IoP | 2015 |
| S28 | Neural-Network based DDoS Detection using Hadoop | A Neural-Network Based DDoS Detection System Using Hadoop And HBase | Teng Zhao | International Symposium on Cyberspace safety and security | 2015 |
| S29 | RADISH | Detecting Insider Threats Using RADISH: A System for Real-Time Anomaly Detection in Heterogeneous Data Streams | Brock B¨ose, Bhargav Avasarala, Srikanta Tirthapura, Yung-Yu Chung, and Donald Steiner | IEEE Systems | 2017 |
| S30 | Forensic Analyzer | Cloud Computing-Based Forensic Analysis for Collaborative Network Security Management System | Zhen Chen, Fuye Han, Junwei Cao, Xin Jiang, and Shuo Chen | Journal of Tsinghua Science and Technology | 2013 |
| S31 | IDS-MRCPSO | MapReduce Intrusion Detection System based on a Particle Swarm Optimization Clustering Algorithm | Ibrahim Aljarah and Simone A. Ludwig | Congress on Evolutionary Computation | 2013 |
| S32 | Hive vs MySQL | Performance Evaluation of Big Data Technology on Designing Big Network Traffic Data Analysis System | Nattawat Khamphakdee, Nunnapus Benjamas and Saiyan Saiyod | Conference on Soft Computing and Intelligent Systems | 2016 |
| S33 | SDN DDoS Detector | A DDoS Detection and Mitigation System Framework Based on Spark and SDN | Qiao Yan(B) and Wenyao Huang | Conference on Smart Computing and Communication | 2017 |
| S34 | Cloud Bursting | High-performance network traffic analysis for continuous batch intrusion detection | Ricardo Morla · Pedro Gonçalves, Jorge G. Barbosa | Journal of supercomputing | 2016 |
| S35 | GPGPU-based IDS | Design Consideration of Network Intrusion Detection System using Hadoop and GPGPU | Sanraj Rajendra Bandre, Jyoti N. Nandimath | Conference on Pervasive Computing | 2015 |
| S36 | BotCloud | BotCloud: Detecting Botnets Using MapReduce | J´erˆome Franc¸ois, Shaonan Wang, Walter Bronzi, Radu State, Thomas Engel | Conference on Information Forensics and Security | 2011 |
| S37 | Ctracer | Ctracer: Uncover C&C in Advanced Persistent Threats based on Scalable Framework for Enterprise Log Data | Kai-Fong Hong, Chien-Chih Chen, Yu-Ting Chiu, and Kuo-Sen Chou | Congress on Big Data | 2015 |
| S38 | SEAS-MR | Scalable Security Event Aggregation for Situation Analysis | Jinoh Kim, Ilhwan Moon, Kyungil Lee, Sang C. Suh, Ikkyun Kim | Conference on Big Data Computing Service and Applications | 2015 |
| S39 | APSIS | Toward a Big Data Architecture for Security Events Analytic | Laila Fetjah, Karim Benzidane, Hassan El Alloussi, Othman El Warrak, Said Jai-Andaloussi and Abderrahim Sekkaki | Conference on Cyber Security and Cloud Computing | 2016 |
| S40 | Security Orchestrator | Enabling Trustworthy Spaces via Orchestrated Analytical Security | Joshua Howes, James Solderitsch, Ignatius Chen, Jonté Craighead | Workshop on Cyber Security and Information Intelligence Research | 2013 |
| S41 | Hunting Attacks in the Dark | Hunting attacks in the dark: clustering and correlation analysis for unsupervised anomaly detection | Johan Mazel, Pedro Casas, Romain Fontugne, Kensuke Fukuda and Philippe Owezarski | Journal of Network Management | 2015 |
| S42 | Multi-Tier Threat Intelligence | Towards a Big Data Architecture for Facilitating Cyber Threat Intelligence | Charles Wheelus, Elias Bou-Harb, Xingquan Zhu | Conference on New Technologies, Mobility, and security | 2016 |
| S43 | DDoS Detection with HTTP Packet Pattern | A method of DDoS attack detection using HTTP packet pattern and rule engine in cloud computing environment | Junho Choi · Chang Choi, Byeongkyu Ko · Pankoo Kim | Journal of soft computing | 2014 |
| S44 | Dynamic Rule Creation based Anomaly Detector | A Dynamic Rule Creation Based Anomaly Detection Method for Identifying Security Breaches in Log Records | Jakub Breier, Jana Braniˇsova | Journal of Wireless Personal Communication | 2017 |
| S45 | VALKYRIE | Valkyrie: Behavioural malware detection using global kernel-level telemetry data | Sven Krasser, Brett Meyer, Patrick Crenshaw | Workshop on Machine learning for signal processing | 2015 |
| S46 | Multiple Detection Methods | Towards online anomaly detection by combining multiple detection methods and Storm | Ziyu Wang, Jiahai Yang, Hui Zhang, Chenxi Li, Shize Zhang and Hui Wang | Conference on Network Operations and Management Symposium | 2016 |
| S47 | APT Detector | Study And Research of APT Detection Technology Based on Big Data Processing Architecture | Lin Shenwen, Li Yingbo, Du Xiongjie | Conference on Electronics Information and Emergency Communication | 2015 |

| ID | Name | Title | Authors | Venue | Year |
|---|---|---|---|---|---|
| S48 | IDS Log Analyzer | Scalable Intrusion Detection Systems Log Analysis using Cloud Computing Infrastructure | Manish Kumar, Dr. M. Hanumanthappa | Conference on Computational Intelligence and Computing Research | 2013 |
| S49 | Data Intensive Framework | Real-Time Handling of Network Monitoring Data Using a Data-Intensive Framework | Aryan Taheri Monfared, Tomasz Wiktor Wlodarczyk, Chunming Rong | Conference on Cloud Computing Technology and Science | 2013 |
| S50 | PhishStorm | PhishStorm: Detecting Phishing With Streaming Analytics | Samuel Marchal, Jérôme François, Radu State, and Thomas Engel | IEEE Transactions on Network and Service Management | 2014 |
| S51 | Lightweight Security Framework | Massive Distributed and Parallel Log Analysis For Organizational Security | Xiaokui Shu, John Smiy, Danfeng (Daphne) Yao, and Heshan Lin | Workshop on security and privacy in big data | 2013 |
| S53 | IoT Monitor | Parallel Processing of Big Heterogeneous Data for Security Monitoring of IoT Networks | Igor Saenko, Igor Kotenko, Alexey Kushnerevich | Conference on Parallel, Distributed, and network-based processing | 2017 |
| S53 | VAST | Native Actors: How to Scale Network Forensics | Matthias Vallentin, Dominik Charousset | SIGCOMM | 2014 |
| S54 | Hybrid Stream and Batch Analyzer | Scalable Hybrid Stream and Hadoop Network Analysis System | Vernon K. C. Bumgardner, Victor W. Marek | Conference on Performance Engineering | 2014 |
| S55 | Web IDS using Storm | Research of Recognition System of Web Intrusion Detection Based on Storm | Li Bo, Wang Jinzhen, Zhao Ping, Yan Zhongjiang, Yang Mao | Conference on Network, communication, and computing | 2016 |
| S56 | Count Me In | Count Me In: Viable Distributed Summary Statistics for Securing High-Speed Networks | Johanna Amann, Seth Hall, and Robin Sommer | Journal on research in attacks, intrusions, and defences | 2014 |
| S57 | PSO Clustering | Towards a Scalable Intrusion Detection System based on Parallel PSO Clustering Using MapReduce | Ibrahim Aljarah and Simone A. Ludwig | Conference on genetic and evolutionary computing | 2013 |
| S58 | Dynamic Time Threshold | A Statistical Threat Detection Method Based on Dynamic Time Threshold | Jian-Wei Tian, Hong Qiao, Xi Li, Zhen Tian | Conference on Computer and Communications | 2016 |
| S59 | Compound Session based Anomaly Detector | A Real-time Network Traffic Anomaly Detection System based on Storm | Gang He, Cheng Tan, Dechen Yu, Xiaochun Wu | Conference on Intelligent Human-Machine Systems and Cybernetics | 2015 |
| S60 | System Call based Detector | A System Call Analysis Method with MapReduce for Malware Detection | Shun-Te Liu, Hui-ching Huang, Yi-Ming Chen | Conference on Parallel and distributed systems | 2011 |
| S61 | Genetic Algo-based DDoS Detector | Distributed Denial of Services Attack Protection System with Genetic Algorithms on Hadoop Cluster Computing Framework | Masataka Mizukoshi | Congress on Evolutionary computation | 2014 |
| S62 | Neural Network & Spark based DDoS Detector | Detection of DDoS attacks based on neural network using apche spark | Chang-Jung Hsieh, Ting-Yuan Chan | Conference on Applied System Innovation | 2016 |
| S63 | Feature Extractor | Distributed Network Traffic Feature Extraction for a Real-time IDS | Ahmad M Karimi, Quamar Niyaz, Weiqing Sun, Ahmad Y Javaid, and Vijay K Devabhaktuni | Conference on Electro Information Technology | 2016 |
| S64 | Hashdoop | Hashdoop: A MapReduce Framework for Network Anomaly Detection | Romain Fontugne, Johan Mazel, Kensuke Fukuda | Workshop on Security and Privacy in Big Data | 2014 |
| S65 | ICAS | ICAS: An Inter-VM IDS Log Cloud Analysis System | Shun-Fa Yang, Wei-Yu Chen, Yao-Tsung Wang | Conf. on Cloud computing and intelligence systems | 2011 |
| S66 | Batch & Stream Analyzer | Real-Time Network Anomaly Detection System Using Machine Learning | Shuai Zhao, Mayanka Chandrashekar, Yugyung Lee, Deep Medhi | Conference on the design of reliable communication networks | 2015 |
| S67 | R-based Traffic Analyzer | Traffic Analysis using Hadoop Cloud | Aishwarya.K, Dr.Sharmila Sankar | Conference on Innovation in Information, Embedded, and communication systems | 2015 |
| S68 | MALSPOT | MalSpot: Multi2 Malicious Network Behavior Patterns Analysis | Hing-Hao Mao1, Chung-Jung Wu1, Evangelos E. Papalexakis, Christos Faloutsos, Kuo-Chen Lee1, and Tien-Cheu | Conference on Knowledge discovery and data mining | 2014 |
| S69 | P2P Bot Detector | Scalable P2P bot detection system based on network data stream | Shree Garg & Sateesh K. Peddoju & Anil K. Sarje | Journal of Peer-to-peer networking applications | 2016 |
| S70 | Hadoop-based Traffic Monitor | Towards Scalable Internet Traffic Measurement and Analysis with Hadoop | Yeonhee Lee and Youngseok Lee | ACM SIGCOMM Computer Communication Review | 2013 |
| S71 | Reliable Traffic Analyzer | An Internet Traffic Analysis Method with MapReduce | Youngseok Lee, Wonchul Kang, Hyeongu Son | Workshop on Network Operations and Management | 2010 |
| S72 | Hybrid IDS | Design of a Hybrid Intrusion Detection System using Snort and Hadoop | Prathibha.P.G, Dileesh.E.D | Journal of Computer Applications | 2013 |
| S73 | PKI-based IDS | Building Scalable Distributed Intrusion Detection Systems Based on the MapReduce Framework | Rafael Timoteo de Sousa Junior | Journal of Revista telecommunications | 2011 |
| S74 | MATATABI | MATATABI: Multi-layer Threat Analysis Platform with Hadoop | Hajime Tazaki, Kazuya Okada, Yuji Sekiya, Youki Kadobayashi | Workshop on Building analysis datasets and gathering experience returns for security | 2014 |

# References

1. Craigen, D., N. Diakun-Thibault, and R. Purse, *Defining cybersecurity.* Technology Innovation Management Review, 2014. **4**(10).
2. Accenture, *Cost of Cyber Crime Study. Available at https://www.accenture.com/t20170926T072837Z__w__/us-en/_acnmedia/PDF-61/Accenture-2017-CostCyberCrimeStudy.pdf.* 2017.
3. Cardenas, A.A., P.K. Manadhata, and S.P. Rajan, *Big data analytics for security.* IEEE Security & Privacy, 2013. **11**(6): p. 74-76.
4. Cárdenas, A.A., P.K. Manadhata, and S. Rajan, *Big data analytics for security intelligence. Available at https://downloads.cloudsecurityalliance.org/initiatives/bdwg/Big_Data_Analytics_for_Security_Intelligence.pdf.* University of Texas at Dallas@ Cloud Security Alliance, 2013: p. 1-22.
5. Chen, C.P. and C.-Y. Zhang, *Data-intensive applications, challenges, techniques and technologies: A survey on Big Data.* Information Sciences, 2014. **275**: p. 314-347.
6. Verma, R., et al., *Security analytics: essential data analytics knowledge for cybersecurity professionals and students.* IEEE Security & Privacy, 2015. **13**(6): p. 60-65.
7. Otero, C.E. and A. Peter, *Research directions for engineering big data analytics software.* IEEE Intelligent Systems, 2015. **30**(1): p. 13-19.
8. Madhavji, N.H., A. Miranskyy, and K. Kontogiannis. *Big picture of big data software engineering: with example research challenges*. in *Big Data Software Engineering (BIGDSE), 2015 IEEE/ACM 1st International Workshop on*. 2015. IEEE.
9. Bass, L., P.C. Clements, and R. Kazman, *Software architecture in practice.* Addison-Wesley, 2012. **3rd ed.**
10. Lewis, G. and P. Lago, *Architectural tactics for cyber-foraging: Results of a systematic literature review.* Journal of Systems and Software, 2015. **107**: p. 158-186.
11. Procaccianti, G., P. Lago, and S. Bevini, *A systematic literature review on energy efficiency in cloud software architectures.* Sustainable Computing: Informatics and Systems, 2015. **7**: p. 2-10.
12. Garcés, L., et al., *Quality attributes and quality models for ambient assisted living software systems: A systematic mapping.* Information and Software Technology, 2017. **82**: p. 121-138.
13. Mahdavi-Hezavehi, S., et al., *A systematic literature review on methods that handle multiple quality attributes in architecture-based self-adaptive systems.* Information and Software Technology, 2017.
14. Kitchenham, B. and S. Charters, *Guidelines for performing systematic literature reviews in software engineering*, in *Technical report, Ver. 2.3 EBSE Technical Report. EBSE*. 2007, sn.
15. Zhang, H., M.A. Babar, and P. Tell, *Identifying relevant studies in software engineering.* Information and Software Technology, 2011. **53**(6): p. 625-637.
16. Wohlin, C. *Guidelines for snowballing in systematic literature studies and a replication in software engineering*. in *Proceedings of the 18th international conference on evaluation and assessment in software engineering*. 2014. ACM.
17. Badampudi, D., C. Wohlin, and K. Petersen. *Experiences from using snowballing and database searches in systematic literature studies*. in *Proceedings of the 19th International Conference on Evaluation and Assessment in Software Engineering*. 2015. ACM.
18. Maplesden, D., et al., *Performance analysis for object-oriented software: A systematic mapping.* IEEE Transactions on Software Engineering, 2015. **41**(7): p. 691-710.
19. Daneva, M., et al., *Empirical research methodologies and studies in Requirements Engineering: How far did we come?* Journal of systems and software, 2014. **95**: p. 1-9.
20. Kitchenham, B., et al., *Systematic literature reviews in software engineering–a tertiary study.* Information and Software Technology, 2010. **52**(8): p. 792-805.
21. Chen, L., M. Ali Babar, and H. Zhang, *Towards an evidence-based understanding of electronic data sources.* 2010.
22. Braun, V. and V. Clarke, *Using thematic analysis in psychology.* Qualitative research in psychology, 2006. **3**(2): p. 77-101.
23. Cruzes, D.S. and T. Dyba. *Recommended steps for thematic synthesis in software engineering*. in *Empirical Software Engineering and Measurement (ESEM), 2011 International Symposium on*. 2011. IEEE.
24. Maier, G., et al. *Enriching network security analysis with time travel*. in *ACM SIGCOMM Computer Communication Review*. 2008. ACM.
25. Park, K., G. Kim, and M. Crovella. *On the relationship between file sizes, transport protocols, and self-similar network traffic*. in *Network Protocols, 1996. Proceedings., 1996 International Conference on*. 1996. IEEE.
26. Paxson, V. and S. Floyd, *Wide area traffic: the failure of Poisson modeling.* IEEE/ACM Transactions on Networking (ToN), 1995. **3**(3): p. 226-244.
27. Pratyusa, A.A.C. and S. Rajan, *Big Data Analytics for Security Intelligence.* CLOUD SECURITY ALLIANCE, 2013.
28. Tan, J., et al. *Improving reducetask data locality for sequential mapreduce jobs*. in *INFOCOM, 2013 Proceedings IEEE*. 2013. IEEE.
29. NetFort, *Flow Analysis Versus Packet Analysis. What Should You Choose? Available at https://www.netfort.com/wp-content/uploads/PDF/WhitePapers/NetFlow-Vs-Packet-Analysis-What-Should-You-Choose.pdf.* White Paper, 2014.

30. Vasilomanolakis, E., et al., *Taxonomy and survey of collaborative intrusion detection.* ACM Computing Surveys (CSUR), 2015. **47**(4): p. 55.
31. Liao, H.-J., et al., *Intrusion detection system: A comprehensive review.* Journal of Network and Computer Applications, 2013. **36**(1): p. 16-24.
32. Mitchell, R. and I.-R. Chen, *A survey of intrusion detection techniques for cyber-physical systems.* ACM Computing Surveys (CSUR), 2014. **46**(4): p. 55.
33. Wiegers, K. and J. Beatty, *Software Requirements.* Microsoft Press, 2014. **Third edition**.
34. Losavio, F., et al., *Quality characteristics for software architecture.* Journal of Object Technology, 2003. **2**(2): p. 133-150.
35. ITRC, *Breach Report. Available at http://www.idtheftcenter.org/Press-Releases/2017-mid-year-data-breach-report-press-release.* 2017.
36. Dreger, H., et al. *Operational experiences with high-volume network intrusion detection.* in *Proceedings of the 11th ACM conference on Computer and communications security.* 2004. ACM.
37. Wikipedia, *Operation Shady RAT. Available at https://en.wikipedia.org/wiki/Operation_Shady_RAT.* 2017.
38. Cisco, *NetFlow. Available at https://pliki.ip-sa.pl/wiki/Wiki.jsp?page=NetFlow. [Last Accessed: 25 Oct, 2017].* 2017.
39. Wong, P.C., et al., *The top 10 challenges in extreme-scale visual analytics.* IEEE computer graphics and applications, 2012. **32**(4): p. 63-67.
40. Zhou, C.V., C. Leckie, and S. Karunasekera, *A survey of coordinated attacks and collaborative intrusion detection.* Computers & Security, 2010. **29**(1): p. 124-140.
41. Stevanovic, M. and J.M. Pedersen, *Machine learning for identifying botnet network traffic.* 2013.
42. Pearson, S. *Taking account of privacy when designing cloud computing services.* in *Proceedings of the 2009 ICSE Workshop on Software Engineering Challenges of Cloud Computing.* 2009. IEEE Computer Society.
43. Ullah, F., et al., *Data Exfiltration: A Review of External Attack Vectors and Countermeasures.* Journal of Network and Computer Applications, 2017.
44. Commission, E., *Protection of Personal Data. Available at http://ec.europa.eu/justice/data-protection/.* 2016.
45. Pfleeger, C.P. and S.L. Pfleeger, *Security in computing.* 2002: Prentice Hall Professional Technical Reference.
46. Zhao, S., et al. *I-can-mama: Integrated campus network monitoring and management.* in *Network Operations and Management Symposium (NOMS), 2014 IEEE.* 2014. IEEE.
47. Frey, B.J. and D. Dueck, *Clustering by passing messages between data points.* science, 2007. **315**(5814): p. 972-976.
48. Cisco, *Cisco Visual Networking Index: Forecast and Methodology, 2016–2021. Available at https://www.cisco.com/c/en/us/solutions/collateral/service-provider/visual-networking-index-vni/complete-white-paper-c11-481360.html.* White Paper, 2017.
49. Bay, S.D., et al., *The UCI KDD archive of large data sets for data mining research and experimentation.* ACM SIGKDD Explorations Newsletter, 2000. **2**(2): p. 81-85.
50. CAIDA, *CAIDA Data. Available at http://www.caida.org/data/.* 2016.
51. Combs, G., *TShark—Dump and Analyze Network Traffic.* Wireshark, 2012.
52. Dean, J. and S. Ghemawat, *MapReduce: simplified data processing on large clusters.* Communications of the ACM, 2008. **51**(1): p. 107-113.
53. Buczak, A.L. and E. Guven, *A survey of data mining and machine learning methods for cyber security intrusion detection.* IEEE Communications Surveys & Tutorials, 2016. **18**(2): p. 1153-1176.
54. Caruana, R. and A. Niculescu-Mizil. *An empirical comparison of supervised learning algorithms.* in *Proceedings of the 23rd international conference on Machine learning.* 2006. ACM.
55. Cheng, C., W.P. Tay, and G.-B. Huang. *Extreme learning machines for intrusion detection.* in *Neural networks (IJCNN), the 2012 international joint conference on.* 2012. IEEE.
56. Qiao, L.-B., et al. *Mining of attack models in ids alerts from network backbone by a two-stage clustering method.* in *Parallel and Distributed Processing Symposium Workshops & PhD Forum (IPDPSW), 2012 IEEE 26th International.* 2012. IEEE.
57. Perdisci, R., G. Giacinto, and F. Roli, *Alarm clustering for intrusion detection systems in computer networks.* Engineering Applications of Artificial Intelligence, 2006. **19**(4): p. 429-438.
58. Sadoddin, R. and A. Ghorbani. *Alert correlation survey: framework and techniques.* in *Proceedings of the 2006 international conference on privacy, security and trust: bridge the gap between PST technologies and business services.* 2006. ACM.
59. Ning, P., Y. Cui, and D.S. Reeves. *Constructing attack scenarios through correlation of intrusion alerts.* in *Proceedings of the 9th ACM Conference on Computer and Communications Security.* 2002. ACM.
60. Aydın, M.A., A.H. Zaim, and K.G. Ceylan, *A hybrid intrusion detection system design for computer network security.* Computers & Electrical Engineering, 2009. **35**(3): p. 517-526.
61. Kabiri, P. and A.A. Ghorbani, *Research on intrusion detection and response: A survey.* IJ Network Security, 2005. **1**(2): p. 84-102.
62. Nieves, J.F. and Y.C. Jiao, *Data clustering for anomaly detection in network intrusion detection.* Research Alliance in Math and Science, 2009: p. 1-12.

63. MIT, *DARPA intrusion detection evaluation data set. Available at http://www.ll.mit.edu/IST/ideval/data/1999/1999_data_index.html* 1999.
64. Kanungo, T., et al., *An efficient k-means clustering algorithm: Analysis and implementation.* IEEE transactions on pattern analysis and machine intelligence, 2002. **24**(7): p. 881-892.
65. Bezdek, J.C., R. Ehrlich, and W. Full, *FCM: The fuzzy c-means clustering algorithm.* Computers & Geosciences, 1984. **10**(2-3): p. 191-203.
66. Arlitt, M., D. Krishnamurthy, and J. Rolia, *Characterizing the scalability of a large web-based shopping system.* ACM Transactions on Internet Technology, 2001. **1**(1): p. 44-69.
67. Challenger, J.R., et al., *Efficiently serving dynamic data at highly accessed web sites.* IEEE/ACM transactions on Networking, 2004. **12**(2): p. 233-246.
68. Kalogeraki, V., P. Melliar-Smith, and L.E. Moser. *Dynamic migration algorithms for distributed object systems*. in *Distributed Computing Systems, 2001. 21st International Conference on.* 2001. IEEE.
69. Lewis, S., et al., *Hydra: a scalable proteomic search engine which utilizes the Hadoop distributed computing framework.* BMC bioinformatics, 2012. **13**(1): p. 324.
70. Pratt, B., et al., *MR-tandem: parallel X! tandem using hadoop MapReduce on amazon Web services.* Bioinformatics, 2011. **28**(1): p. 136-137.
71. Wiley, K., et al. *Astronomical image processing with hadoop*. in *Astronomical Data Analysis Software and Systems XX*. 2011.
72. Loebman, S., et al. *Analyzing massive astrophysical datasets: Can Pig/Hadoop or a relational DBMS help?* in *Cluster Computing and Workshops, 2009. CLUSTER'09. IEEE International Conference on*. 2009. IEEE.
73. Zhang, F., et al., *A task-level adaptive MapReduce framework for real-time streaming data in healthcare applications.* Future Generation Computer Systems, 2015. **43**: p. 149-160.
74. White, T., *Hadoop: The Definitive Guide. O'Reilly.* Scbastopol, California, 2009.
75. Lee, K.-H., et al., *Parallel data processing with MapReduce: a survey.* AcM sIGMoD Record, 2012. **40**(4): p. 11-20.
76. Lee, Y., W. Kang, and Y. Lee. *A hadoop-based packet trace processing tool*. in *International Workshop on Traffic Monitoring and Analysis*. 2011. Springer.
77. Aljarah, I. and S.A. Ludwig. *Parallel particle swarm optimization clustering algorithm based on mapreduce methodology*. in *Nature and biologically inspired computing (NaBIC), 2012 fourth world congress on*. 2012. IEEE.
78. White, T., *Hadoop: The Definitive Guide. Availalbe at http://javaarm.com/file/apache/Hadoop/books/Hadoop-The.Definitive.Guide_4.edition_a_Tom.White_April-2015.pdf*. O'Reilly, 2015.
79. Apache, *Flume. Available at http://flume.apache.org/. [Last Accessed: 1st Nov, 20127]*. 2017.
80. Cisco, *Cisco Systems NetFlow Services Export Version 9. Available at https://www.ietf.org/rfc/rfc3954.txt. [Last Accessed: 24 Oct, 2017]*.
81. Vinoski, S., *Advanced message queuing protocol.* IEEE Internet Computing, 2006. **10**(6).
82. Wikipedia, *Link Aggregation. Available at https://en.wikipedia.org/wiki/Link_aggregation. [Last Accessed 24 Oct, 2017]*. 2016.
83. Bass, T., *Intrusion detection systems and multisensor data fusion.* Communications of the ACM, 2000. **43**(4): p. 99-105.
84. Muttik, I. and C. Barton, *Cloud security technologies.* Information security technical report, 2009. **14**(1): p. 1-6.
85. Myers, M., et al., *X. 509 Internet public key infrastructure online certificate status protocol-OCSP*. 1999.
86. Vaarandi, R. *Real-time classification of IDS alerts with data mining techniques*. in *Military Communications Conference, 2009. MILCOM 2009. IEEE*. 2009. IEEE.
87. Viinikka, J., et al., *Processing intrusion detection alert aggregates with time series modeling.* Information Fusion, 2009. **10**(4): p. 312-324.
88. Julisch, K., *Clustering intrusion detection alarms to support root cause analysis.* ACM transactions on information and system security (TISSEC), 2003. **6**(4): p. 443-471.
89. Allier, S., et al. *A framework to compare alert ranking algorithms*. in *Reverse Engineering (WCRE), 2012 19th Working Conference on*. 2012. IEEE.
90. Anderson, R.J., *Security engineering: a guide to building dependable distributed systems*. 2010: John Wiley & Sons.
91. Bass, L., *Software architecture in practice*. 2007: Pearson Education India.
92. Gorton, I., *Essential software architecture*. 2006: Springer Science & Business Media.
93. Dyba, T., B.A. Kitchenham, and M. Jorgensen, *Evidence-based software engineering for practitioners.* IEEE software, 2005. **22**(1): p. 58-65.
94. Devanbu, P., T. Zimmermann, and C. Bird. *Belief & evidence in empirical software engineering*. in *Proceedings of the 38th international conference on software engineering*. 2016. ACM.
95. Grahn, K., M. Westerlund, and G. Pulkkis, *Analytics for Network Security: A Survey and Taxonomy*, in *Information Fusion for Cyber-Security Analytics*. 2017, Springer. p. 175-193.
96. Wang, L. and R. Jones, *Big Data Analytics for Network Intrusion Detection: A Survey.* International Journal of Networks and Communications, 2017. **7**(1): p. 24-31.
97. Mahmood, T. and U. Afzal. *Security Analytics: Big Data Analytics for cybersecurity: A review of trends, techniques and tools*. in *Information assurance (ncia), 2013 2nd national conference on*. 2013. IEEE.

98. Jeong, H.-D.J., et al. *Anomaly teletraffic intrusion detection systems on hadoop-based platforms: A survey of some problems and solutions*. in *Network-Based Information Systems (NBiS), 2012 15th International Conference on*. 2012. IEEE.
99. Alguliyev, R. and Y. Imamverdiyev. *Big data: big promises for information security*. in *Application of Information and Communication Technologies (AICT), 2014 IEEE 8th International Conference on*. 2014. IEEE.
100. Zuech, R., T.M. Khoshgoftaar, and R. Wald, *Intrusion detection and big heterogeneous data: a survey.* Journal of Big Data, 2015. **2**(1): p. 3.
101. Dyba, T., T. Dingsoyr, and G.K. Hanssen. *Applying systematic reviews to diverse study types: An experience report*. in *Empirical Software Engineering and Measurement, 2007. ESEM 2007. First International Symposium on*. 2007. IEEE.